\documentclass[aps,prl,reprint,superscriptaddress,nobibnotes]{revtex4-1}

\usepackage{textcomp}
\usepackage{upgreek}
\usepackage{setspace}
\usepackage{esint}
\usepackage{array}

\newcommand{\te}[1]{\textrm{#1}}
\usepackage{graphicx}

\begin{document}


\title{Spin-orbit interaction in a dual gated InAs/GaSb quantum well}


\author{Arjan J. A. Beukman}
\author{Folkert K. de Vries}
\author{Jasper van Veen}
\author{Rafal Skolasinski}
\author{Michael Wimmer}
\author{Fanming Qu}
\author{David T. de Vries}

\affiliation{QuTech and Kavli Institute of Nanoscience, Delft University of Technology, 2600 GA Delft, The Netherlands}

\author{Binh-Minh Nguyen}
\author{Wei Yi}
\author{Andrey A. Kiselev}
\author{Marko Sokolich}
\affiliation{HRL Laboratories, 3011 Malibu Canyon Road, Malibu, California 90265, USA}

\author{Michael J. Manfra}
\affiliation{Department of Physics and Astronomy and Station Q Purdue, Purdue University, West Lafayette, Indiana 47907, USA}

\author{Fabrizio Nichele}
\author{Charles M. Marcus}
\affiliation{Center for Quantum Devices, Niels Bohr Institute, University of Copenhagen, 2100 Copenhagen, Denmark}

\author{Leo P. Kouwenhoven}
\email{l.p.kouwenhoven@tudelft.nl}
\affiliation{QuTech and Kavli Institute of Nanoscience, Delft University of Technology, 2600 GA Delft, The Netherlands}


\date{\today}

\begin{abstract}
Spin-orbit interaction is investigated in a dual gated InAs/GaSb quantum well. Using an electric field the quantum well can be tuned between a single carrier regime with exclusively electrons as carriers and a two-carriers regime where electrons and holes coexist. 
Spin-orbit interaction in both regimes manifests itself as a beating in the Shubnikov-de Haas oscillations. 
In the single carrier regime the linear Dresselhaus strength is characterized by $\beta = 28.5$ meV\AA \ and the Rashba coefficient $\alpha$ is tuned from 75 to 53 meV\AA \ by changing the electric field. In the two-carriers regime the spin splitting shows a nonmonotonic behavior with gate voltage, which is consistent with our band structure calculations.

\end{abstract}


\maketitle

The semiconductors InAs and GaSb have small band gaps together with a crystal inversion asymmetry resulting from their zincblende structure. These materials are therefore predicted to have strong spin-orbit interaction (SOI) \cite{Winkler_2003, Fabian_2007} which has been measured experimentally \cite{Nitta_1997}. Moreover, tuning of the Rashba strength by electrostatic gating has been shown for InAs quantum wells \cite{Grundler_2000,Shojaei_2016}. 
Strong and in-situ control over SOI is a necessary ingredient for novel spintronic devices \cite{Zutic_2004, Fabian_2007, Datta_1990}, and strong SOI together with a large g-factor and induced superconductivity are ingredients for a topological superconducting phase \cite{Alicea_2012}.

\begin{figure}[h!t]
	\centering
	\includegraphics[width=65mm]{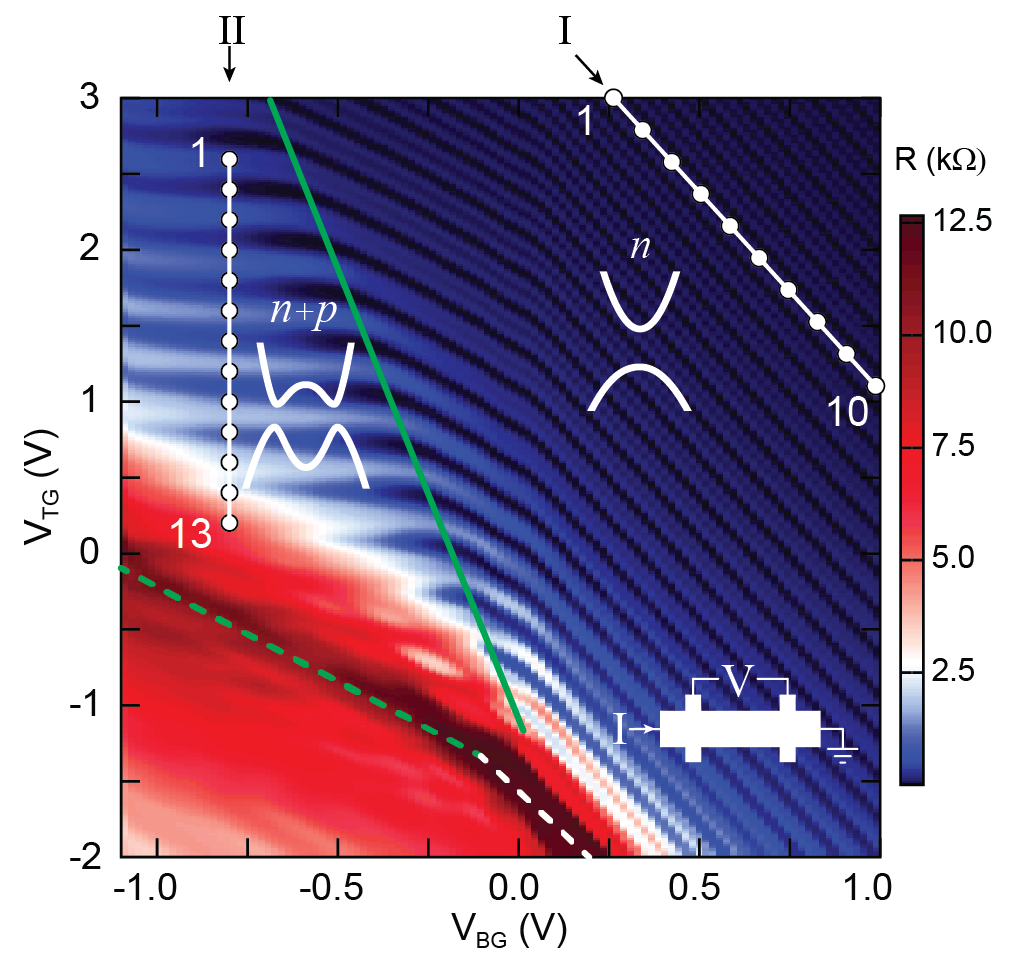}
	\caption{Longitudinal resistance of the Hall bar device (see bottom right inset) as a function of top gate voltage ($V_{tg}$) and back gate voltage ($V_{bg}$) at 2 T out of plane magnetic field. Oscillations in resistance originate from Landau levels and denote lines of constant electron density. 
	The dashed green and white lines indicate regions with the Fermi level located inside an energy gap. The solid green line separates the region with electrons as carriers (right) from a region where electrons and holes coexist (left). 
	Line I is situated in the electron regime and  Line II in the two-carrier regime. The insets show the schematic band alignment for both cases.
	}
	\label{SO:fig1}
\end{figure}

Combining InAs and GaSb in a quantum well gained much interest because of the type-II broken-gap band alignment \cite{Kroemer_2004}.
As a result, the GaSb valence band maximum is higher in energy than the InAs conduction band minimum, opening a range of energies where electrons in the InAs coexist with holes in the GaSb. 
The spatial separation of these electron and hole gases allows for tunability of the band alignment using an electric field. Therefore, a rich phase diagram can be mapped out using dual gated devices \cite{Liu_2008, Qu_2015}.
Although spatially separated, strong coupling between the materials allows for electron-hole hybridization which opens a gap in the energy spectrum when the density of electrons equals that of holes \cite{AndradaeSilva_1994, Cooper_1998}, driving the band structure topologically non-trivial \cite{Liu_2008}.

Interestingly, the magnitude of this hybridization gap is spin dependent due to SOI \cite{Zakharova_2001, Halvorsen_2000, Xu_2010}. 
Therefore, a spin polarized state is seen at energies close to the hybridization gap \cite{Nichele_2016} and at higher energies a dip in the spin-splitting is expected \cite{Li_2008}. The latter has yet to be observed. 
Here, we experimentally study SOI through the difference in density of the spin-orbit split bands of an InAs/GaSb quantum well. This zero-field density difference ($\Delta n_{\tiny \textit{ZF}}$) is extracted from magnetoresistance measurements.
First, SOI is investigated in the regime where the GaSb is depleted from carriers. Rashba and Dresselhaus SOI strengths can be extracted from measurements of $\Delta n_{\tiny \textit{ZF}}$.
Second, SOI is investigated just above the hybridization gap where $\Delta n_{\tiny \textit{ZF}}$ almost vanishes, consistent with band structure calculations.

A 20 \textmu m wide and 80 \textmu m long Hall bar device is defined using chemical wet etching techniques. 
A top gate is separated from the mesa by a 80 nm thick SiN$_\textrm{\scriptsize x}$ dielectric layer. 
The Hall bar is fabricated from the same wafer used in Ref. \cite{Nguyen2015, Qu_2015}. The quantum well consists of 12.5 nm InAs and 5 nm GaSb between 50 nm AlSb barriers. The doped GaSb substrate acts as a back gate. 
All measurements are done at 300 mK using standard lock-in techniques with an excitation current of 50 nA.

Figure 1 presents the longitudinal resistance of the Hall bar device as a function of top gate voltage $V_\mathrm{tg}$ and back gate voltage $V_\mathrm{bg}$. 
The measurement is performed in 2 T perpendicular magnetic field and therefore shows quantum oscillations resulting from the changing electron density. 
Quantum oscillations corresponding to holes are less pronounced as the mobility of holes in this system is much lower than the mobility of electrons \cite{Qu_2015}. 
For lines parallel to these oscillations, such as line I in Fig. 1a, the electron density is constant while the electric field changes. 
Regions of high resistance, indicated by the dashed white and green lines, correspond to having the Fermi level inside an energy gap. 
A detailed description of the phase diagram obtained from measurements on the same wafer was reported by Qu et al. \cite{Qu_2015}.

The green solid line in Fig. 1 divides the phase diagram in two regimes. Right of this line is the electron-only regime, where the GaSb is depleted. 
The system effectively is an asymmetric InAs quantum well with a trivial band alignment and a Fermi level residing in the conduction band (see inset of Fig. 1a). 
In this regime we investigate $\Delta n_{\tiny \textit{ZF}}$ along line I, where the electron mobility is highest while only the lowest subband remains occupied. 
The regime at the left of the green line is the two-carriers regime where electrons and holes coexist. 
Line II is chosen to evaluate $\Delta n_{\tiny \textit{ZF}}$ close to the hybridization gap (highlighted by the dashed green line). 
Before discussing the spin-orbit interaction in the two-carriers regime (along line II) we first study the electron-only regime (line I).

\begin{figure}[h]
	\centering
	\includegraphics{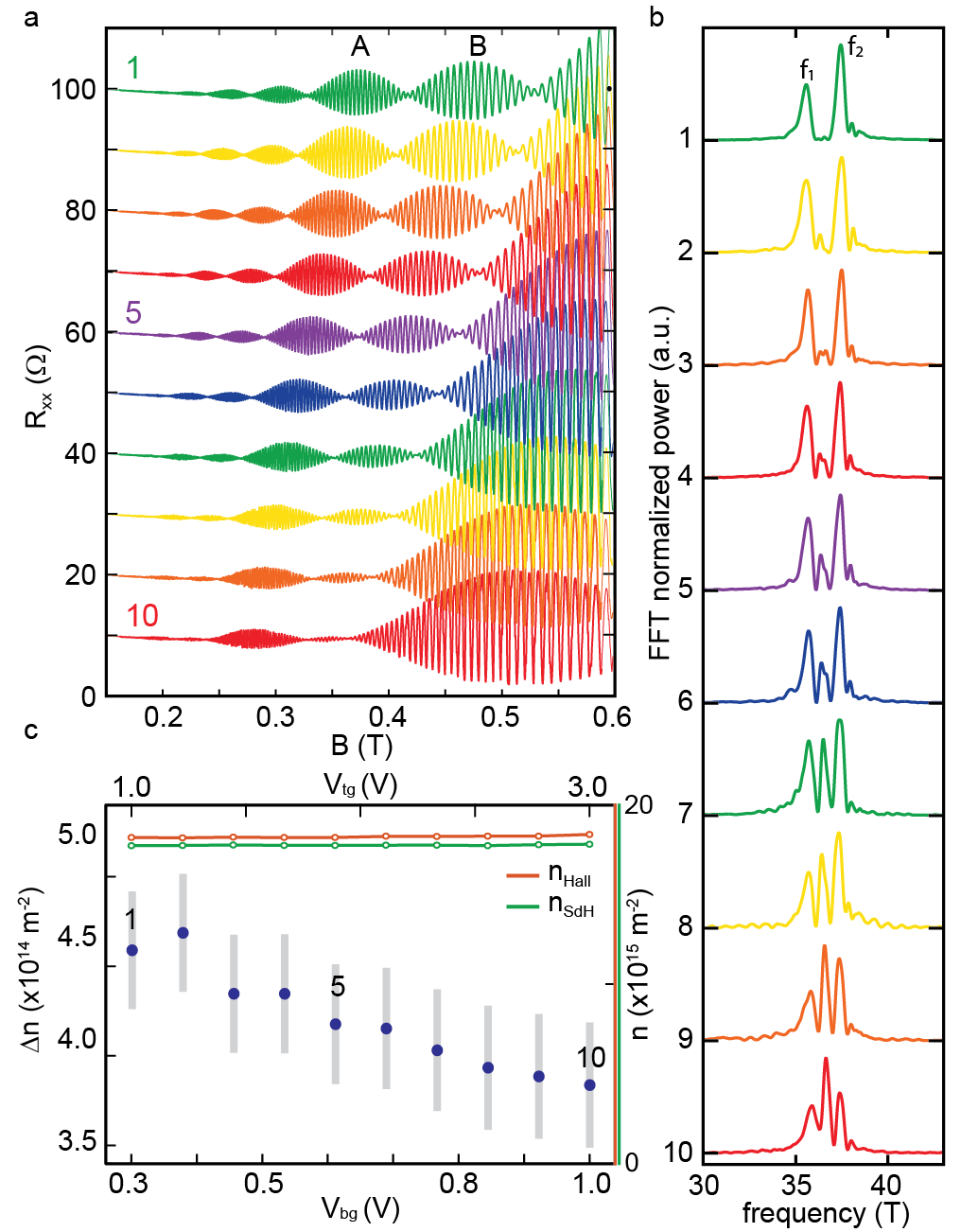}
	\caption{Spin-splitting at a constant electron density in the electron regime. (a) Magnetoresistance traces for data points 1-10 along line I indicated in Fig. 1. 
	A constant background is subtracted from the traces and they are offset 10 $\Omega$ from each other. 
	(b) Fourier power spectra $|\mathcal{F}[R_{xx}(1/B)]|^2$ of the traces in (a). 
	(c) Electron density extracted from Hall resistance and Shubnikov-de Haas period (right axis) together with the $\Delta n_{\tiny \textit{ZF}}$ at each data point along line I (left axis), with error bars in gray.
		}
	\label{SO:fig2}
\end{figure}

Figure 2a shows magnetoresistance traces for 10 points along line I. 
The density of electrons is fixed (see Fig. 2c) while the electric field is changed. We first consider trace 1. 
Clear oscillations in the longitudinal resistance $R_{xx}$ are observed as a function of perpendicular magnetic field $B$ modulated by a beat pattern. 
These Shubnikov-de Haas (SdH) oscillations appear for each single spin band and are periodic in $1/B$ with a frequency that relates to the carrier density via $n=e/h \cdot f$ \cite{Nitta_1997, Onsager_1952}. 
The beat modulation observed in trace 1 is caused by two slightly different SdH frequencies $f_1,f_2$. 
This is also evident from the fast Fourier transform (FFT) of the magnetoresistance trace $\mathcal{F}[R_{xx}(1/B)]$ presented in the first curve of Fig. 2b (see supplementary info for details on the Fourier procedure \cite{Suppl}). 
These two SdH frequencies indicate two distinct densities $n_1,n_2$. 
They must correspond to different spin species because their sum $n_1+n_2$ equals the Hall density $n_H$ (see Fig. 2c). Subsequently, one spin species has a larger density than the other, $n_1>n_2$, implying that the system favors one spin-orbit eigenstate over to the other. The difference, $\Delta n_{\tiny \textit{ZF}} = n_2-n_1$, is a measure for the zero-field spin splitting energy, $\Delta E_{\scriptsize \textrm{ZFSS}} = \Delta n_{\tiny \textit{ZF}} \left( m^*/\pi \hbar^2 \right)^{-1} $. 

Upon moving from point 1 to 10 along line I, two trends are observed. First, an extra frequency peak emerges in the FFTs at $(f_1+f_2)/2$. 
This originates from the asymmetry between adjacent beats in the SdH oscillations, visible both in amplitude and number of oscillations of beats A and B in Fig. 2a \cite{Suppl}. 
Second, the spacing between the outer peaks in the FFT spectrum decreases, evident from decreasing $\Delta n_{\tiny \textit{ZF}}$ over line I (Fig. 2c). 
This arises from an increasing number of oscillations in both beats A and B (see \cite{Suppl}), which also pushes the beat nodes to lower magnetic fields. 
Before we extract the actual SOI strengths and show its electric field dependence, we first elucidate the origin of the emerging center frequency peak. 

The center frequency, interestingly, does not correspond to an actual density. 
The sum of the densities $n_1$ and $n_2$ (corresponding to the outer peaks in the FFT) still equals the Hall density. There are, however, mechanisms involving scattering between Fermi-surfaces that can result in extra frequency components. 
Such mechanisms are magnetic inter subband scattering (MIS) \cite{Sander_1998, Rowe_2001}, magnetophonon resonances (MPR) \cite{Tsui_1980, Gurevich_1961} and magnetic breakdown (MB) \cite{Averkiev_2005, Symons_1998, Shoenberg_1984}. 

We exclude MIS and MPR. By changing electron density all the frequency peak positions shift with equal strength \cite{Suppl}. 
However, the oscillation frequency of MIS and MPR is determined by the subband spacing and a specific phonon frequency, respectively. 
Both do not depend on the electron density. 
In contrast, for MB the spurious peak always appears in between $f_1$ and $f_2$. 
Magnetic breakdown explains this spurious central peak as carriers tunneling between spin polarized Fermi-surfaces at spin-degeneracy points. The interplay of Dresselhaus and Rashba SOI in our heterostructure could lead to such an anisotropic Fermi surface \cite{Averkiev_2005, Ganichev_2004}. 
In order to confirm this hypothesis, we extract the individual Rashba and Dresselhaus contributions by comparing our data to quantum mechanical Landau level simulations that include the MB mechanism. 

The quantum well in this electron-only regime is modeled by a Hamiltonian with spin-orbit interaction in 2D electron systems subject to a perpendicular magnetic field $B$, as given by \cite{Winkler_2003, Fabian_2007}: 
\begin{widetext}
\begin{equation}
	H = \frac{(\hat{p}_x^2 + \hat{p}_y^2)}{2m^*} \sigma_0 + \alpha(\hat{p}_y \sigma_x-\hat{p}_x\sigma_y)/\hbar + \beta (\hat{p}_x\sigma_x - \hat{p}_y\sigma_y)/\hbar + \gamma (\hat{p}_y \hat{p}_x \hat{p}_y \sigma_x - \hat{p}_x \hat{p}_y \hat{p}_x \sigma_y)/\hbar^3 + \frac{1}{2}g\mu_B B \sigma_z
\end{equation}
\end{widetext}
Where $p_i \rightarrow p_i + eA_i$ is the canonical momentum, $\sigma_i$ Pauli spin matrices, $\alpha, \beta, \gamma$ the Rashba, linear Dresselhaus and cubic Dresselhaus coefficients, respectively, $\hbar$ the reduced Planck’s constant, $\mu_B$ the Bohr magneton. 
An electron effective mass of $m^*=0.04 m_0$ is measured from the temperature dependence of the SdH oscillations \cite{Suppl} and a g-factor of $-11.5$ is used in the calculations \cite{Mu_2016, footnote_one}.
We solve for the Landau level energies in a perpendicular magnetic field $B_z$ and extract the resistivity as a function of magnetic field (see supplementary information for details \cite{Suppl}).

\begin{figure}[h]
	\centering
	\includegraphics[width=86mm]{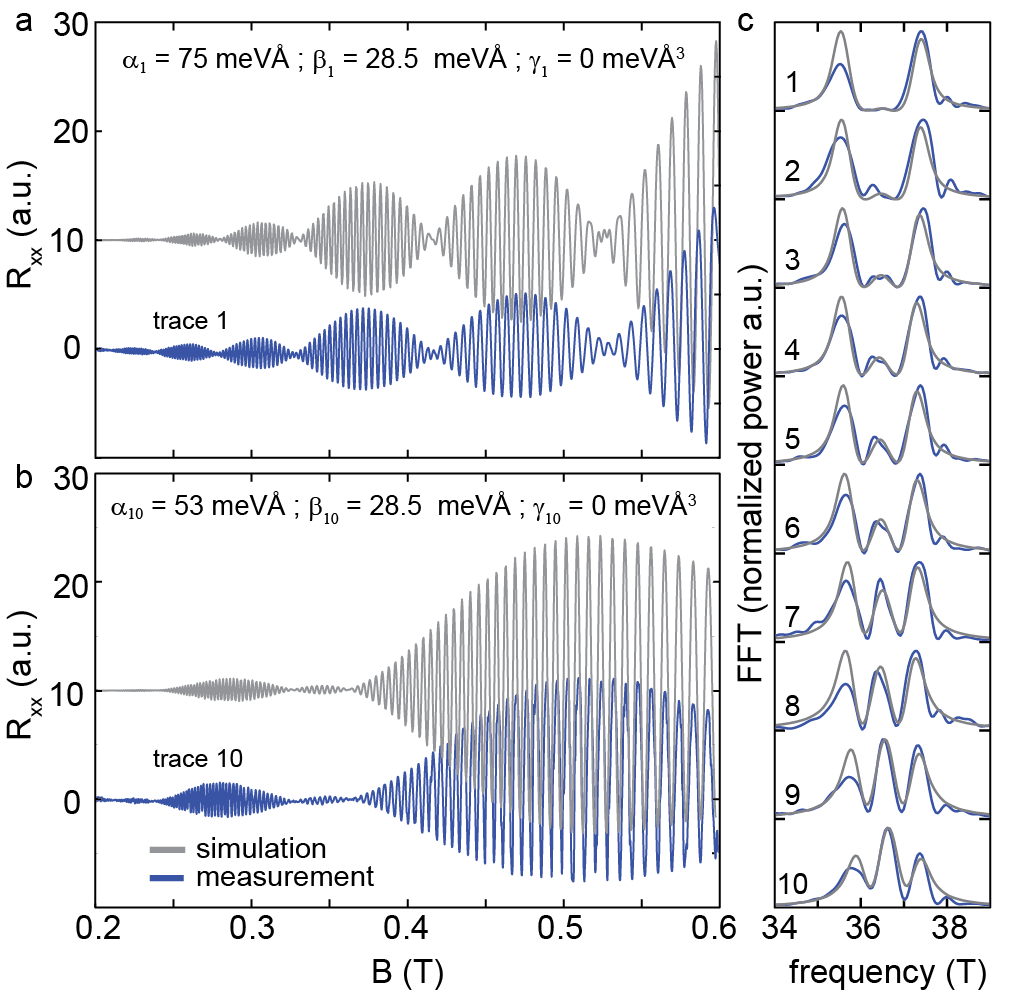}
	\caption{Landau level simulations for a 2DEG with Rashba and Dresselhaus spin-orbit interaction. 
	(a) and (b) depict the measured trace (blue) together with the simulated magnetoresistance trace (gray) which is offset by 10 units. 
	The values for $\alpha, \beta,$ and  $\gamma$ used are mentioned in the figure. In all the simulations the Landau level broadening is set to $\Gamma = 0.45$ meV. 
	(c) Fast Fourier transform of the simulated and measured magnetoresistance at points 1 to 10 along line I. All simulated magnetoresistance traces can be found in the supplementary information \cite{Suppl}. }
	\label{SO:fig3}
\end{figure}

The parameters $\alpha, \beta, \gamma$ in the model are estimated and fine tuned to match the node positions and the number of oscillations in a beat of the measured SdH traces. 
Figure 3 a,b shows the measured SdH data together with the simulated data for traces 1 and 10. Trace 1 is fitted with $\alpha_1 = 75$ meV\AA , $\beta_1=28.5$ meV\AA , $\gamma_1=0$ meV\AA$^3$ and trace 10 is fitted with $\alpha_{10}=53$ meV\AA \ , $\beta_{10}=28.5$ meV\AA \ ,$\gamma_{10}=0$ meV\AA$^3$. 
The node positions and amplitude modulation of the simulated data agrees well with the measured SdH oscillations. 

Curiously, only good fits are obtained when setting the cubic Dresselhaus term $\gamma$ to zero. 
In 2D systems, $\beta$ is related to $\gamma$ via $\beta=\langle k_z^2 \rangle \gamma$, where $\langle k_z^2 \rangle \approx (\pi/d)^2$ is the expectation value of the transverse momentum \cite{Winkler_2003, Fabian_2007} in a quantum well of thickness $d$. 
So $\gamma$ should be non-zero.
Currently we do not understand this discrepancy. A recent experimental study on a similar material system also found that the cubic Dresselhaus term could be neglected \cite{Herzog_2017}.

Now we consider all traces (1-10) and show that the two trends of Fig. 2 (emerging center FFT peak and approaching outer FFT peaks) are reproduced by changing only the Rashba SOI strength.
Figure 3c shows the FFTs of the simulated traces where $\alpha$ is linearly interpolated between $\alpha_1$ and $\alpha_{10}$ while fixing $\beta=28.5$ meV\AA \ and $\gamma=0$ meV\AA$^3$. 
Linear interpolation is used because the electric field changes linearly along line I, and Rashba SOI strength depends linearly on electric field \cite{Bychkov_1984a, Bychkov_1984b, Winkler_2003}. 
All simulated FFTs and the SdH traces \cite{Suppl} match the measured data very well, clearly reproducing the emerging central peak and the approaching outer peaks.

In the remainder of this paper we switch to the two-carriers regime, located left of the solid green line in Fig. 1. 
Electrons in InAs are present alongside with holes in GaSb (n+p). 
Here we study the influence of the hybridization of electrons with holes on $\Delta n_{\tiny \textit{ZF}}$ by investigating magnetoresistance traces on the points 1-13 along line II. 

Before continuing with the measured magnetoresistance traces, it is insightful to examine the expected band structures at points 1 and 13, as illustrated in Fig. 4b. 
The first point of line II is located near the boundary between the two-carrier and single carrier regimes. 
A small amount of holes with a large amount of electrons is present. At point 13, close to the hybridization gap, the electron and hole densities are roughly equal, hence the Fermi level $E_f$ is close to the hybridization gap. 
Note also that  $k_\textrm{\scriptsize cross}$ decreases from 1 to 13, since the electric field changes.

\begin{figure}[h]
	\centering
	\includegraphics{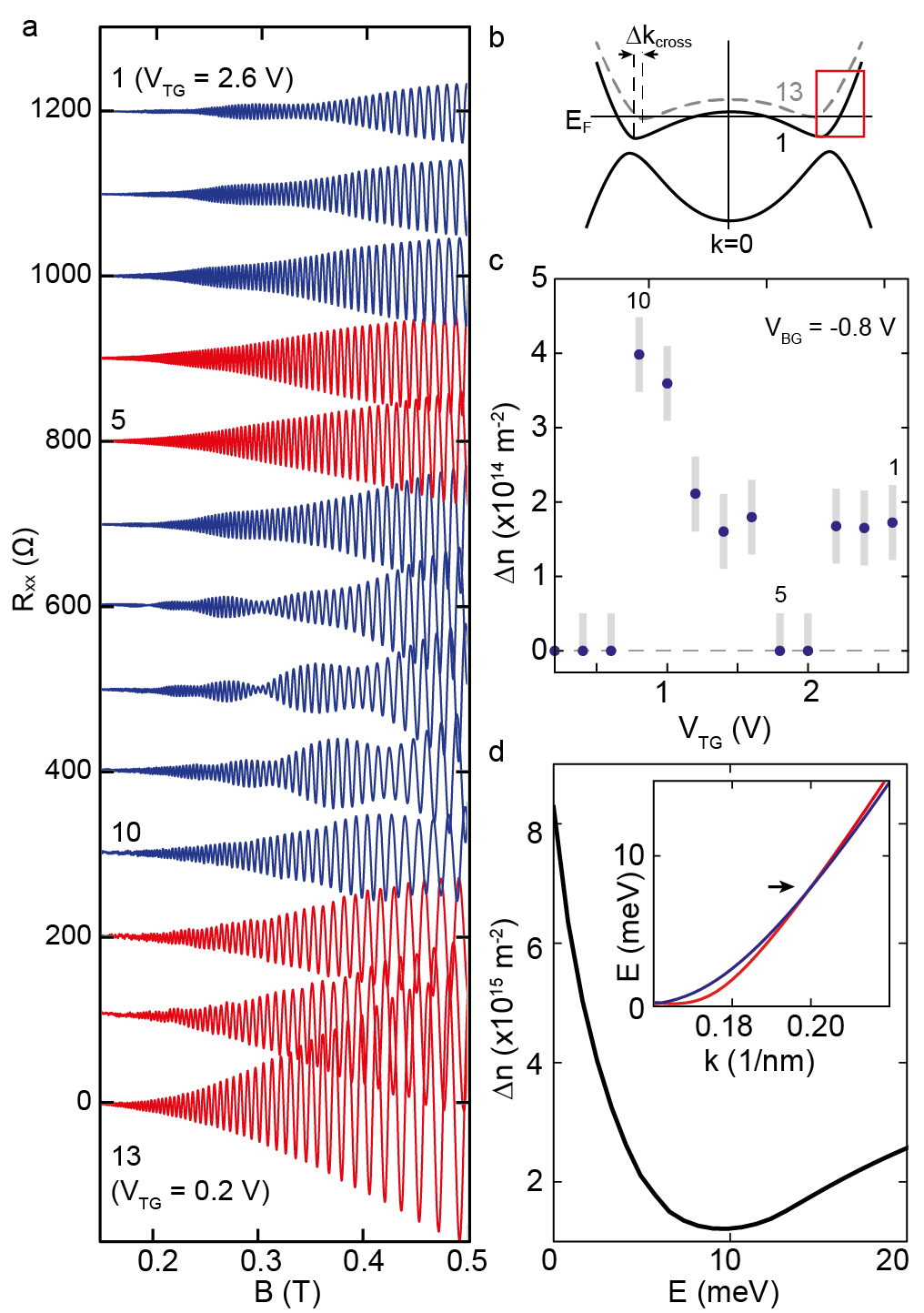}
	\caption{Spin-splitting in the two-carrier regime.
	(a) Magnetoresistance traces for points 1 to 13 along line II indicated in Fig. 1. 
	For each trace the $R_{xx}(B=0)$ background resistance is subtracted and afterwards the traces are offset by 100 $\Omega$. 
	Beating is (not) observed for (red) blue colored traces. 
	(b) Schematic band structure tuning when moving from point 1 to 13. 
	(c) $\Delta n_{\tiny \textit{ZF}}$ extracted from the Fourier transform of magnetoresistance traces of (a). Error bars are indicated by the light blue bar. 
	(d) $\Delta n_{\tiny \textit{ZF}}$ extracted from band structure calculation for our InAs/GaSb quantum well at zero electric field. The inset shows the corresponding band structure in the [100] direction.
	}
	\label{SO:fig4}
\end{figure}

Figure 4a shows the magnetoresistance traces 1-13 along line II. 
Starting from trace 1 towards trace 13 we find series of traces with or without beating, depicted in blue and red respectively.
For traces 1 to 3, at large electron density, beating is observed from which we extract $\Delta n_{\tiny \textit{ZF}} = 1.7 \cdot 10^{14}$ m$^{-2}$ \footnote{We cannot directly extract the spin-orbit strength from this $\Delta n$ by comparing to the single-carrier case, since the effective mass in this region is unknown.}. 
Remarkably trace 4 and 5 do not show any beating, therefore no zero-field density difference can be extracted. 
For traces 6 to 10, the beating revives showing strong beating. Finally, traces 11-13 show no beating. 
Figure 4c depicts the extracted $\Delta n_{\tiny \textit{ZF}}$ along line II, which shows a non-monotonic behaviour as a function of gate voltage along line II.

In order to understand this non-monotonic $\Delta n_{\tiny \textit{ZF}}$ near the hybridization gap (points 1-10) we performed band structure calculations of our InAs/GaSb quantum well \cite{Suppl}.
The $\Delta n$ extracted from these calculations is plotted in Fig. 4d, which qualitatively agree with the observed dip in $\Delta n_{\tiny \textit{ZF}}$ at points 4 and 5 (Fig. 4c.). In order to understand the simulated $\Delta n$, the band structure near the hybridization gap is depicted in the inset of Fig. 4d (zoom-in on Fig. 4b indicated by the red box). 
The black and red lines represent different spin bands. 
The bands cross at the black arrow, indicating the vanishing of $\Delta n$, such as observed in the experiment. 
We found this feature to be robust for different electric fields and crystal directions \cite{Suppl}. We therefore attribute the observed dip in $\Delta n_{\tiny \textit{ZF}}$ to the band structure near the hybridization gap.

Note that only qualitative comparison between experiment and calculations is possible as Fermi-energy is varied in the simulation, while in the experiment the band structure ($k_\textrm{\scriptsize cross}$) and Fermi-energy are expected to change. The fact that $\Delta n_{\tiny \textit{ZF}}$ in Fig. 4d does not completely vanish is because the crossing of the spin bands in the [110] occurs at a slightly different energy than in the [100] direction.

The lack of beating of traces 11-13 is not captured with the simulation. There are two possible reasons for this deviation. First, a strong asymmetry in SdH amplitudes of the two spin species ($A_\textrm{\scriptsize up} \gg A_\textrm{\scriptsize down}$) determines the visibility of the beating pattern.
The single spin band SdH oscillation amplitude depends on effective mass $m^*$ and scattering time according to $A_\textrm{\scriptsize SdH} \sim (eB/m^*)^3 \exp(-\pi /\omega_c \tau)$ \cite{Luo_1990}. Both effective mass and scattering time for the two spin bands become very dissimilar when approaching the hybridization gap \cite{Suppl}, as a result that the beating visibility is reduced to below the experimentally detectable visibility.
Second, Nichele et al. \cite{Nichele_2016} shows there is an energy window with only one single spin band present. 
In such spin polarized state no beating can occur. 
Here, we cannot discriminate between these two reasons that explain the lack of beating in traces 11-13. 

In conclusion, we presented a study of the spin-orbit interaction in an InAs/GaSb double quantum well. 
The Fermi-level and band structure are altered by top and bottom gates. In the electron-only regime we find a electric field tunable spin-orbit interaction, and extract the individual Rashba and Dresselhaus terms.
In the two-carriers regime we observe a non-monotonic behavior of the spin splitting which we trace back to the crossing of the spin bands due to the hybridization of electrons and holes.

\begin{acknowledgments}
We gratefully acknowledge Roland Winkler for very helpful discussions. This work has been supported by funding from the Netherlands
Foundation for Fundamental Research on Matter (FOM) and
Microsoft Corporation Station Q.
\end{acknowledgments}

\bibliography{so_paper_arxiv_version}

\begin{thebibliography}{36}%
\makeatletter
\providecommand \@ifxundefined [1]{%
 \@ifx{#1\undefined}
}%
\providecommand \@ifnum [1]{%
 \ifnum #1\expandafter \@firstoftwo
 \else \expandafter \@secondoftwo
 \fi
}%
\providecommand \@ifx [1]{%
 \ifx #1\expandafter \@firstoftwo
 \else \expandafter \@secondoftwo
 \fi
}%
\providecommand \natexlab [1]{#1}%
\providecommand \enquote  [1]{``#1''}%
\providecommand \bibnamefont  [1]{#1}%
\providecommand \bibfnamefont [1]{#1}%
\providecommand \citenamefont [1]{#1}%
\providecommand \href@noop [0]{\@secondoftwo}%
\providecommand \href [0]{\begingroup \@sanitize@url \@href}%
\providecommand \@href[1]{\@@startlink{#1}\@@href}%
\providecommand \@@href[1]{\endgroup#1\@@endlink}%
\providecommand \@sanitize@url [0]{\catcode `\\12\catcode `\$12\catcode
  `\&12\catcode `\#12\catcode `\^12\catcode `\_12\catcode `\%12\relax}%
\providecommand \@@startlink[1]{}%
\providecommand \@@endlink[0]{}%
\providecommand \url  [0]{\begingroup\@sanitize@url \@url }%
\providecommand \@url [1]{\endgroup\@href {#1}{\urlprefix }}%
\providecommand \urlprefix  [0]{URL }%
\providecommand \Eprint [0]{\href }%
\providecommand \doibase [0]{http://dx.doi.org/}%
\providecommand \selectlanguage [0]{\@gobble}%
\providecommand \bibinfo  [0]{\@secondoftwo}%
\providecommand \bibfield  [0]{\@secondoftwo}%
\providecommand \translation [1]{[#1]}%
\providecommand \BibitemOpen [0]{}%
\providecommand \bibitemStop [0]{}%
\providecommand \bibitemNoStop [0]{.\EOS\space}%
\providecommand \EOS [0]{\spacefactor3000\relax}%
\providecommand \BibitemShut  [1]{\csname bibitem#1\endcsname}%
\let\auto@bib@innerbib\@empty
\bibitem [{\citenamefont {Winkler}(2003)}]{Winkler_2003}%
  \BibitemOpen
  \bibfield  {author} {\bibinfo {author} {\bibfnamefont {R.}~\bibnamefont
  {Winkler}},\ }\href
  {http://www.ebook.de/de/product/2638263/roland_winkler_spin_orbit_coupling_effects_in_two_dimensional_electron_and_hole_systems.html}
  {\emph {\bibinfo {title} {Spin-orbit Coupling Effects in Two-dimensional
  Electron and Hole Systems}}}\ (\bibinfo  {publisher} {Springer-Verlag GmbH},\
  \bibinfo {year} {2003})\BibitemShut {NoStop}%
\bibitem [{\citenamefont {Fabian}\ \emph {et~al.}(2007)\citenamefont {Fabian},
  \citenamefont {Matos-Abiague}, \citenamefont {Ertler}, \citenamefont
  {Stano},\ and\ \citenamefont {{\v{Z}}uti{\'{c}}}}]{Fabian_2007}%
  \BibitemOpen
  \bibfield  {author} {\bibinfo {author} {\bibfnamefont {J.}~\bibnamefont
  {Fabian}}, \bibinfo {author} {\bibfnamefont {A.}~\bibnamefont
  {Matos-Abiague}}, \bibinfo {author} {\bibfnamefont {C.}~\bibnamefont
  {Ertler}}, \bibinfo {author} {\bibfnamefont {P.}~\bibnamefont {Stano}}, \
  and\ \bibinfo {author} {\bibfnamefont {I.}~\bibnamefont
  {{\v{Z}}uti{\'{c}}}},\ }\href {\doibase 10.2478/v10155-010-0086-8} {\bibfield
   {journal} {\bibinfo  {journal} {Acta Physica Slovaca. Reviews and
  Tutorials}\ }\textbf {\bibinfo {volume} {57}} (\bibinfo {year} {2007}),\
  10.2478/v10155-010-0086-8}\BibitemShut {NoStop}%
\bibitem [{\citenamefont {Nitta}\ \emph {et~al.}(1997)\citenamefont {Nitta},
  \citenamefont {Akazaki}, \citenamefont {Takayanagi},\ and\ \citenamefont
  {Enoki}}]{Nitta_1997}%
  \BibitemOpen
  \bibfield  {author} {\bibinfo {author} {\bibfnamefont {J.}~\bibnamefont
  {Nitta}}, \bibinfo {author} {\bibfnamefont {T.}~\bibnamefont {Akazaki}},
  \bibinfo {author} {\bibfnamefont {H.}~\bibnamefont {Takayanagi}}, \ and\
  \bibinfo {author} {\bibfnamefont {T.}~\bibnamefont {Enoki}},\ }\href
  {\doibase 10.1103/physrevlett.78.1335} {\bibfield  {journal} {\bibinfo
  {journal} {Phys. Rev. Lett.}\ }\textbf {\bibinfo {volume} {78}},\ \bibinfo
  {pages} {1335} (\bibinfo {year} {1997})}\BibitemShut {NoStop}%
\bibitem [{\citenamefont {Grundler}(2000)}]{Grundler_2000}%
  \BibitemOpen
  \bibfield  {author} {\bibinfo {author} {\bibfnamefont {D.}~\bibnamefont
  {Grundler}},\ }\href {\doibase 10.1103/physrevlett.84.6074} {\bibfield
  {journal} {\bibinfo  {journal} {Phys. Rev. Lett.}\ }\textbf {\bibinfo
  {volume} {84}},\ \bibinfo {pages} {6074} (\bibinfo {year}
  {2000})}\BibitemShut {NoStop}%
\bibitem [{\citenamefont {Shojaei}\ \emph {et~al.}(2016)\citenamefont
  {Shojaei}, \citenamefont {O{\textquotesingle}Malley}, \citenamefont
  {Shabani}, \citenamefont {Roushan}, \citenamefont {Schultz}, \citenamefont
  {Lutchyn}, \citenamefont {Nayak}, \citenamefont {Martinis},\ and\
  \citenamefont {Palmstr{\o}m}}]{Shojaei_2016}%
  \BibitemOpen
  \bibfield  {author} {\bibinfo {author} {\bibfnamefont {B.}~\bibnamefont
  {Shojaei}}, \bibinfo {author} {\bibfnamefont {P.~J.~J.}\ \bibnamefont
  {O{\textquotesingle}Malley}}, \bibinfo {author} {\bibfnamefont
  {J.}~\bibnamefont {Shabani}}, \bibinfo {author} {\bibfnamefont
  {P.}~\bibnamefont {Roushan}}, \bibinfo {author} {\bibfnamefont {B.~D.}\
  \bibnamefont {Schultz}}, \bibinfo {author} {\bibfnamefont {R.~M.}\
  \bibnamefont {Lutchyn}}, \bibinfo {author} {\bibfnamefont {C.}~\bibnamefont
  {Nayak}}, \bibinfo {author} {\bibfnamefont {J.~M.}\ \bibnamefont {Martinis}},
  \ and\ \bibinfo {author} {\bibfnamefont {C.~J.}\ \bibnamefont
  {Palmstr{\o}m}},\ }\href {\doibase 10.1103/physrevb.93.075302} {\bibfield
  {journal} {\bibinfo  {journal} {Phys. Rev. B}\ }\textbf {\bibinfo {volume}
  {93}} (\bibinfo {year} {2016}),\ 10.1103/physrevb.93.075302}\BibitemShut
  {NoStop}%
\bibitem [{\citenamefont {{\v{Z}}uti{\'{c}}}\ \emph {et~al.}(2004)\citenamefont
  {{\v{Z}}uti{\'{c}}}, \citenamefont {Fabian},\ and\ \citenamefont
  {Sarma}}]{Zutic_2004}%
  \BibitemOpen
  \bibfield  {author} {\bibinfo {author} {\bibfnamefont {I.}~\bibnamefont
  {{\v{Z}}uti{\'{c}}}}, \bibinfo {author} {\bibfnamefont {J.}~\bibnamefont
  {Fabian}}, \ and\ \bibinfo {author} {\bibfnamefont {S.~D.}\ \bibnamefont
  {Sarma}},\ }\href {\doibase 10.1103/revmodphys.76.323} {\bibfield  {journal}
  {\bibinfo  {journal} {Reviews of Modern Physics}\ }\textbf {\bibinfo {volume}
  {76}},\ \bibinfo {pages} {323} (\bibinfo {year} {2004})}\BibitemShut
  {NoStop}%
\bibitem [{\citenamefont {Datta}\ and\ \citenamefont {Das}(1990)}]{Datta_1990}%
  \BibitemOpen
  \bibfield  {author} {\bibinfo {author} {\bibfnamefont {S.}~\bibnamefont
  {Datta}}\ and\ \bibinfo {author} {\bibfnamefont {B.}~\bibnamefont {Das}},\
  }\href {\doibase 10.1063/1.102730} {\bibfield  {journal} {\bibinfo  {journal}
  {Appl. Phys. Lett.}\ }\textbf {\bibinfo {volume} {56}},\ \bibinfo {pages}
  {665} (\bibinfo {year} {1990})}\BibitemShut {NoStop}%
\bibitem [{\citenamefont {Alicea}(2012)}]{Alicea_2012}%
  \BibitemOpen
  \bibfield  {author} {\bibinfo {author} {\bibfnamefont {J.}~\bibnamefont
  {Alicea}},\ }\href {\doibase 10.1088/0034-4885/75/7/076501} {\bibfield
  {journal} {\bibinfo  {journal} {Rep. Prog. Phys.}\ }\textbf {\bibinfo
  {volume} {75}},\ \bibinfo {pages} {076501} (\bibinfo {year}
  {2012})}\BibitemShut {NoStop}%
\bibitem [{\citenamefont {Kroemer}(2004)}]{Kroemer_2004}%
  \BibitemOpen
  \bibfield  {author} {\bibinfo {author} {\bibfnamefont {H.}~\bibnamefont
  {Kroemer}},\ }\href {\doibase 10.1016/j.physe.2003.08.003} {\bibfield
  {journal} {\bibinfo  {journal} {Physica E: Low-dimensional Systems and
  Nanostructures}\ }\textbf {\bibinfo {volume} {20}},\ \bibinfo {pages} {196}
  (\bibinfo {year} {2004})}\BibitemShut {NoStop}%
\bibitem [{\citenamefont {Liu}\ \emph {et~al.}(2008)\citenamefont {Liu},
  \citenamefont {Hughes}, \citenamefont {Qi}, \citenamefont {Wang},\ and\
  \citenamefont {Zhang}}]{Liu_2008}%
  \BibitemOpen
  \bibfield  {author} {\bibinfo {author} {\bibfnamefont {C.}~\bibnamefont
  {Liu}}, \bibinfo {author} {\bibfnamefont {T.~L.}\ \bibnamefont {Hughes}},
  \bibinfo {author} {\bibfnamefont {X.-L.}\ \bibnamefont {Qi}}, \bibinfo
  {author} {\bibfnamefont {K.}~\bibnamefont {Wang}}, \ and\ \bibinfo {author}
  {\bibfnamefont {S.-C.}\ \bibnamefont {Zhang}},\ }\href {\doibase
  10.1103/physrevlett.100.236601} {\bibfield  {journal} {\bibinfo  {journal}
  {Phys. Rev. Lett.}\ }\textbf {\bibinfo {volume} {100}} (\bibinfo {year}
  {2008}),\ 10.1103/physrevlett.100.236601}\BibitemShut {NoStop}%
\bibitem [{\citenamefont {Qu}\ \emph {et~al.}(2015)\citenamefont {Qu},
  \citenamefont {Beukman}, \citenamefont {Nadj-Perge}, \citenamefont {Wimmer},
  \citenamefont {Nguyen}, \citenamefont {Yi}, \citenamefont {Thorp},
  \citenamefont {Sokolich}, \citenamefont {Kiselev}, \citenamefont {Manfra},
  \citenamefont {Marcus},\ and\ \citenamefont {Kouwenhoven}}]{Qu_2015}%
  \BibitemOpen
  \bibfield  {author} {\bibinfo {author} {\bibfnamefont {F.}~\bibnamefont
  {Qu}}, \bibinfo {author} {\bibfnamefont {A.~J.~A.}\ \bibnamefont {Beukman}},
  \bibinfo {author} {\bibfnamefont {S.}~\bibnamefont {Nadj-Perge}}, \bibinfo
  {author} {\bibfnamefont {M.}~\bibnamefont {Wimmer}}, \bibinfo {author}
  {\bibfnamefont {B.-M.}\ \bibnamefont {Nguyen}}, \bibinfo {author}
  {\bibfnamefont {W.}~\bibnamefont {Yi}}, \bibinfo {author} {\bibfnamefont
  {J.}~\bibnamefont {Thorp}}, \bibinfo {author} {\bibfnamefont
  {M.}~\bibnamefont {Sokolich}}, \bibinfo {author} {\bibfnamefont {A.~A.}\
  \bibnamefont {Kiselev}}, \bibinfo {author} {\bibfnamefont {M.~J.}\
  \bibnamefont {Manfra}}, \bibinfo {author} {\bibfnamefont {C.~M.}\
  \bibnamefont {Marcus}}, \ and\ \bibinfo {author} {\bibfnamefont {L.~P.}\
  \bibnamefont {Kouwenhoven}},\ }\href {\doibase
  10.1103/physrevlett.115.036803} {\bibfield  {journal} {\bibinfo  {journal}
  {Phys. Rev. Lett.}\ }\textbf {\bibinfo {volume} {115}} (\bibinfo {year}
  {2015}),\ 10.1103/physrevlett.115.036803}\BibitemShut {NoStop}%
\bibitem [{\citenamefont {de~Andrada~e Silva}\ \emph
  {et~al.}(1994)\citenamefont {de~Andrada~e Silva}, \citenamefont {Rocca},\
  and\ \citenamefont {Bassani}}]{AndradaeSilva_1994}%
  \BibitemOpen
  \bibfield  {author} {\bibinfo {author} {\bibfnamefont {E.~A.}\ \bibnamefont
  {de~Andrada~e Silva}}, \bibinfo {author} {\bibfnamefont {G.~C.~L.}\
  \bibnamefont {Rocca}}, \ and\ \bibinfo {author} {\bibfnamefont
  {F.}~\bibnamefont {Bassani}},\ }\href {\doibase 10.1103/physrevb.50.8523}
  {\bibfield  {journal} {\bibinfo  {journal} {Phys. Rev. B}\ }\textbf {\bibinfo
  {volume} {50}},\ \bibinfo {pages} {8523} (\bibinfo {year}
  {1994})}\BibitemShut {NoStop}%
\bibitem [{\citenamefont {Cooper}\ \emph {et~al.}(1998)\citenamefont {Cooper},
  \citenamefont {Patel}, \citenamefont {Drouot}, \citenamefont {Linfield},
  \citenamefont {Ritchie},\ and\ \citenamefont {Pepper}}]{Cooper_1998}%
  \BibitemOpen
  \bibfield  {author} {\bibinfo {author} {\bibfnamefont {L.~J.}\ \bibnamefont
  {Cooper}}, \bibinfo {author} {\bibfnamefont {N.~K.}\ \bibnamefont {Patel}},
  \bibinfo {author} {\bibfnamefont {V.}~\bibnamefont {Drouot}}, \bibinfo
  {author} {\bibfnamefont {E.~H.}\ \bibnamefont {Linfield}}, \bibinfo {author}
  {\bibfnamefont {D.~A.}\ \bibnamefont {Ritchie}}, \ and\ \bibinfo {author}
  {\bibfnamefont {M.}~\bibnamefont {Pepper}},\ }\href {\doibase
  10.1103/physrevb.57.11915} {\bibfield  {journal} {\bibinfo  {journal} {Phys.
  Rev. B}\ }\textbf {\bibinfo {volume} {57}},\ \bibinfo {pages} {11915}
  (\bibinfo {year} {1998})}\BibitemShut {NoStop}%
\bibitem [{\citenamefont {Zakharova}\ \emph {et~al.}(2001)\citenamefont
  {Zakharova}, \citenamefont {Yen},\ and\ \citenamefont
  {Chao}}]{Zakharova_2001}%
  \BibitemOpen
  \bibfield  {author} {\bibinfo {author} {\bibfnamefont {A.}~\bibnamefont
  {Zakharova}}, \bibinfo {author} {\bibfnamefont {S.}~\bibnamefont {Yen}}, \
  and\ \bibinfo {author} {\bibfnamefont {K.}~\bibnamefont {Chao}},\ }\href
  {\doibase 10.1103/physrevb.64.235332} {\bibfield  {journal} {\bibinfo
  {journal} {Phys. Rev. B}\ }\textbf {\bibinfo {volume} {64}} (\bibinfo {year}
  {2001}),\ 10.1103/physrevb.64.235332}\BibitemShut {NoStop}%
\bibitem [{\citenamefont {Halvorsen}\ \emph {et~al.}(2000)\citenamefont
  {Halvorsen}, \citenamefont {Galperin},\ and\ \citenamefont
  {Chao}}]{Halvorsen_2000}%
  \BibitemOpen
  \bibfield  {author} {\bibinfo {author} {\bibfnamefont {E.}~\bibnamefont
  {Halvorsen}}, \bibinfo {author} {\bibfnamefont {Y.}~\bibnamefont {Galperin}},
  \ and\ \bibinfo {author} {\bibfnamefont {K.~A.}\ \bibnamefont {Chao}},\
  }\href {\doibase 10.1103/physrevb.61.16743} {\bibfield  {journal} {\bibinfo
  {journal} {Phys. Rev. B}\ }\textbf {\bibinfo {volume} {61}},\ \bibinfo
  {pages} {16743} (\bibinfo {year} {2000})}\BibitemShut {NoStop}%
\bibitem [{\citenamefont {Xu}\ \emph {et~al.}(2010)\citenamefont {Xu},
  \citenamefont {Li}, \citenamefont {Dong}, \citenamefont {Gumbs},\ and\
  \citenamefont {Folkes}}]{Xu_2010}%
  \BibitemOpen
  \bibfield  {author} {\bibinfo {author} {\bibfnamefont {W.}~\bibnamefont
  {Xu}}, \bibinfo {author} {\bibfnamefont {L.~L.}\ \bibnamefont {Li}}, \bibinfo
  {author} {\bibfnamefont {H.~M.}\ \bibnamefont {Dong}}, \bibinfo {author}
  {\bibfnamefont {G.}~\bibnamefont {Gumbs}}, \ and\ \bibinfo {author}
  {\bibfnamefont {P.~A.}\ \bibnamefont {Folkes}},\ }\href {\doibase
  10.1063/1.3476059} {\bibfield  {journal} {\bibinfo  {journal} {J. Appl.
  Phys.}\ }\textbf {\bibinfo {volume} {108}},\ \bibinfo {pages} {053709}
  (\bibinfo {year} {2010})}\BibitemShut {NoStop}%
\bibitem [{\citenamefont {Nichele}\ \emph {et~al.}(2016)\citenamefont
  {Nichele}, \citenamefont {Kjaergaard}, \citenamefont {Suominen},
  \citenamefont {Skolasinski}, \citenamefont {Wimmer}, \citenamefont {Nguyen},
  \citenamefont {Kiselev}, \citenamefont {Yi}, \citenamefont {Sokolich},
  \citenamefont {Manfra}, \citenamefont {Qu}, \citenamefont {Beukman},
  \citenamefont {Kouwenhoven},\ and\ \citenamefont {Marcus}}]{Nichele_2016}%
  \BibitemOpen
  \bibfield  {author} {\bibinfo {author} {\bibfnamefont {F.}~\bibnamefont
  {Nichele}}, \bibinfo {author} {\bibfnamefont {M.}~\bibnamefont {Kjaergaard}},
  \bibinfo {author} {\bibfnamefont {H.~J.}\ \bibnamefont {Suominen}}, \bibinfo
  {author} {\bibfnamefont {R.}~\bibnamefont {Skolasinski}}, \bibinfo {author}
  {\bibfnamefont {M.}~\bibnamefont {Wimmer}}, \bibinfo {author} {\bibfnamefont
  {B.-M.}\ \bibnamefont {Nguyen}}, \bibinfo {author} {\bibfnamefont {A.~A.}\
  \bibnamefont {Kiselev}}, \bibinfo {author} {\bibfnamefont {W.}~\bibnamefont
  {Yi}}, \bibinfo {author} {\bibfnamefont {M.}~\bibnamefont {Sokolich}},
  \bibinfo {author} {\bibfnamefont {M.~J.}\ \bibnamefont {Manfra}}, \bibinfo
  {author} {\bibfnamefont {F.}~\bibnamefont {Qu}}, \bibinfo {author}
  {\bibfnamefont {A.~J.~A.}\ \bibnamefont {Beukman}}, \bibinfo {author}
  {\bibfnamefont {L.~P.}\ \bibnamefont {Kouwenhoven}}, \ and\ \bibinfo {author}
  {\bibfnamefont {C.~M.}\ \bibnamefont {Marcus}},\ }\href@noop {} {\bibfield
  {journal} {\bibinfo  {journal} {ArXiv}\ } (\bibinfo {year} {2016})},\ \Eprint
  {http://arxiv.org/abs/1605.01241} {1605.01241} \BibitemShut {NoStop}%
\bibitem [{\citenamefont {Li}\ \emph {et~al.}(2008)\citenamefont {Li},
  \citenamefont {Chang}, \citenamefont {Hai},\ and\ \citenamefont
  {Chan}}]{Li_2008}%
  \BibitemOpen
  \bibfield  {author} {\bibinfo {author} {\bibfnamefont {J.}~\bibnamefont
  {Li}}, \bibinfo {author} {\bibfnamefont {K.}~\bibnamefont {Chang}}, \bibinfo
  {author} {\bibfnamefont {G.~Q.}\ \bibnamefont {Hai}}, \ and\ \bibinfo
  {author} {\bibfnamefont {K.~S.}\ \bibnamefont {Chan}},\ }\href {\doibase
  10.1063/1.2909544} {\bibfield  {journal} {\bibinfo  {journal} {Appl. Phys.
  Lett.}\ }\textbf {\bibinfo {volume} {92}},\ \bibinfo {pages} {152107}
  (\bibinfo {year} {2008})}\BibitemShut {NoStop}%
\bibitem [{\citenamefont {Nguyen}\ \emph {et~al.}(2015)\citenamefont {Nguyen},
  \citenamefont {Yi}, \citenamefont {Noah}, \citenamefont {Thorp},\ and\
  \citenamefont {Sokolich}}]{Nguyen2015}%
  \BibitemOpen
  \bibfield  {author} {\bibinfo {author} {\bibfnamefont {B.-M.}\ \bibnamefont
  {Nguyen}}, \bibinfo {author} {\bibfnamefont {W.}~\bibnamefont {Yi}}, \bibinfo
  {author} {\bibfnamefont {R.}~\bibnamefont {Noah}}, \bibinfo {author}
  {\bibfnamefont {J.}~\bibnamefont {Thorp}}, \ and\ \bibinfo {author}
  {\bibfnamefont {M.}~\bibnamefont {Sokolich}},\ }\href {\doibase
  10.1063/1.4906589} {\bibfield  {journal} {\bibinfo  {journal} {Applied
  Physics Letters}\ }\textbf {\bibinfo {volume} {106}},\ \bibinfo {pages}
  {032107} (\bibinfo {year} {2015})}\BibitemShut {NoStop}%
\bibitem [{\citenamefont {Onsager}(1952)}]{Onsager_1952}%
  \BibitemOpen
  \bibfield  {author} {\bibinfo {author} {\bibfnamefont {L.}~\bibnamefont
  {Onsager}},\ }\href {\doibase 10.1080/14786440908521019} {\bibfield
  {journal} {\bibinfo  {journal} {The London, Edinburgh, and Dublin
  Philosophical Magazine and Journal of Science}\ }\textbf {\bibinfo {volume}
  {43}},\ \bibinfo {pages} {1006} (\bibinfo {year} {1952})}\BibitemShut
  {NoStop}%
\bibitem [{Sup()}]{Suppl}%
  \BibitemOpen
  \href@noop {} {}\bibinfo {note} {See the supplemental material attached to
  this manuscript.}\BibitemShut {Stop}%
\bibitem [{\citenamefont {Sander}\ \emph {et~al.}(1998)\citenamefont {Sander},
  \citenamefont {Holmes}, \citenamefont {Harris}, \citenamefont {Maude},\ and\
  \citenamefont {Portal}}]{Sander_1998}%
  \BibitemOpen
  \bibfield  {author} {\bibinfo {author} {\bibfnamefont {T.~H.}\ \bibnamefont
  {Sander}}, \bibinfo {author} {\bibfnamefont {S.~N.}\ \bibnamefont {Holmes}},
  \bibinfo {author} {\bibfnamefont {J.~J.}\ \bibnamefont {Harris}}, \bibinfo
  {author} {\bibfnamefont {D.~K.}\ \bibnamefont {Maude}}, \ and\ \bibinfo
  {author} {\bibfnamefont {J.~C.}\ \bibnamefont {Portal}},\ }\href {\doibase
  10.1103/physrevb.58.13856} {\bibfield  {journal} {\bibinfo  {journal} {Phys.
  Rev. B}\ }\textbf {\bibinfo {volume} {58}},\ \bibinfo {pages} {13856}
  (\bibinfo {year} {1998})}\BibitemShut {NoStop}%
\bibitem [{\citenamefont {Rowe}\ \emph {et~al.}(2001)\citenamefont {Rowe},
  \citenamefont {Nehls}, \citenamefont {Stradling},\ and\ \citenamefont
  {Ferguson}}]{Rowe_2001}%
  \BibitemOpen
  \bibfield  {author} {\bibinfo {author} {\bibfnamefont {A.~C.~H.}\
  \bibnamefont {Rowe}}, \bibinfo {author} {\bibfnamefont {J.}~\bibnamefont
  {Nehls}}, \bibinfo {author} {\bibfnamefont {R.~A.}\ \bibnamefont
  {Stradling}}, \ and\ \bibinfo {author} {\bibfnamefont {R.~S.}\ \bibnamefont
  {Ferguson}},\ }\href {\doibase 10.1103/physrevb.63.201307} {\bibfield
  {journal} {\bibinfo  {journal} {Phys. Rev. B}\ }\textbf {\bibinfo {volume}
  {63}} (\bibinfo {year} {2001}),\ 10.1103/physrevb.63.201307}\BibitemShut
  {NoStop}%
\bibitem [{\citenamefont {Tsui}\ \emph {et~al.}(1980)\citenamefont {Tsui},
  \citenamefont {Englert}, \citenamefont {Cho},\ and\ \citenamefont
  {Gossard}}]{Tsui_1980}%
  \BibitemOpen
  \bibfield  {author} {\bibinfo {author} {\bibfnamefont {D.~C.}\ \bibnamefont
  {Tsui}}, \bibinfo {author} {\bibfnamefont {T.}~\bibnamefont {Englert}},
  \bibinfo {author} {\bibfnamefont {A.~Y.}\ \bibnamefont {Cho}}, \ and\
  \bibinfo {author} {\bibfnamefont {A.~C.}\ \bibnamefont {Gossard}},\ }\href
  {\doibase 10.1103/physrevlett.44.341} {\bibfield  {journal} {\bibinfo
  {journal} {Phys. Rev. Lett.}\ }\textbf {\bibinfo {volume} {44}},\ \bibinfo
  {pages} {341} (\bibinfo {year} {1980})}\BibitemShut {NoStop}%
\bibitem [{\citenamefont {Gurevich}\ and\ \citenamefont
  {Firsov}(1961)}]{Gurevich_1961}%
  \BibitemOpen
  \bibfield  {author} {\bibinfo {author} {\bibfnamefont {V.}~\bibnamefont
  {Gurevich}}\ and\ \bibinfo {author} {\bibfnamefont {Y.~A.}\ \bibnamefont
  {Firsov}},\ }\href@noop {} {\bibfield  {journal} {\bibinfo  {journal} {Soviet
  Physics Jetp-Ussr}\ }\textbf {\bibinfo {volume} {13}},\ \bibinfo {pages}
  {137} (\bibinfo {year} {1961})}\BibitemShut {NoStop}%
\bibitem [{\citenamefont {Averkiev}\ \emph {et~al.}(2005)\citenamefont
  {Averkiev}, \citenamefont {Glazov},\ and\ \citenamefont
  {Tarasenko}}]{Averkiev_2005}%
  \BibitemOpen
  \bibfield  {author} {\bibinfo {author} {\bibfnamefont {N.}~\bibnamefont
  {Averkiev}}, \bibinfo {author} {\bibfnamefont {M.}~\bibnamefont {Glazov}}, \
  and\ \bibinfo {author} {\bibfnamefont {S.}~\bibnamefont {Tarasenko}},\ }\href
  {\doibase 10.1016/j.ssc.2004.12.005} {\bibfield  {journal} {\bibinfo
  {journal} {Solid State Communications}\ }\textbf {\bibinfo {volume} {133}},\
  \bibinfo {pages} {543} (\bibinfo {year} {2005})}\BibitemShut {NoStop}%
\bibitem [{\citenamefont {Symons}\ \emph {et~al.}(1998)\citenamefont {Symons},
  \citenamefont {Lakrimi}, \citenamefont {Nicholas}, \citenamefont {Maude},
  \citenamefont {Portal}, \citenamefont {Mason},\ and\ \citenamefont
  {Walker}}]{Symons_1998}%
  \BibitemOpen
  \bibfield  {author} {\bibinfo {author} {\bibfnamefont {D.~M.}\ \bibnamefont
  {Symons}}, \bibinfo {author} {\bibfnamefont {M.}~\bibnamefont {Lakrimi}},
  \bibinfo {author} {\bibfnamefont {R.~J.}\ \bibnamefont {Nicholas}}, \bibinfo
  {author} {\bibfnamefont {D.~K.}\ \bibnamefont {Maude}}, \bibinfo {author}
  {\bibfnamefont {J.~C.}\ \bibnamefont {Portal}}, \bibinfo {author}
  {\bibfnamefont {N.~J.}\ \bibnamefont {Mason}}, \ and\ \bibinfo {author}
  {\bibfnamefont {P.~J.}\ \bibnamefont {Walker}},\ }\href {\doibase
  10.1103/physrevb.58.7292} {\bibfield  {journal} {\bibinfo  {journal} {Phys.
  Rev. B}\ }\textbf {\bibinfo {volume} {58}},\ \bibinfo {pages} {7292}
  (\bibinfo {year} {1998})}\BibitemShut {NoStop}%
\bibitem [{\citenamefont {Shoenberg}(1984)}]{Shoenberg_1984}%
  \BibitemOpen
  \bibfield  {author} {\bibinfo {author} {\bibfnamefont {D.}~\bibnamefont
  {Shoenberg}},\ }\href {\doibase 10.1017/cbo9780511897870} {\emph {\bibinfo
  {title} {Magnetic oscillations in metals}}}\ (\bibinfo  {publisher}
  {Cambridge University Press ({CUP})},\ \bibinfo {year} {1984})\BibitemShut
  {NoStop}%
\bibitem [{\citenamefont {Ganichev}\ \emph {et~al.}(2004)\citenamefont
  {Ganichev}, \citenamefont {Bel'kov}, \citenamefont {Golub}, \citenamefont
  {Ivchenko}, \citenamefont {Schneider}, \citenamefont {Giglberger},
  \citenamefont {Eroms}, \citenamefont {Boeck}, \citenamefont {Borghs},
  \citenamefont {Wegscheider}, \citenamefont {Weiss},\ and\ \citenamefont
  {Prettl}}]{Ganichev_2004}%
  \BibitemOpen
  \bibfield  {author} {\bibinfo {author} {\bibfnamefont {S.~D.}\ \bibnamefont
  {Ganichev}}, \bibinfo {author} {\bibfnamefont {V.~V.}\ \bibnamefont
  {Bel'kov}}, \bibinfo {author} {\bibfnamefont {L.~E.}\ \bibnamefont {Golub}},
  \bibinfo {author} {\bibfnamefont {E.~L.}\ \bibnamefont {Ivchenko}}, \bibinfo
  {author} {\bibfnamefont {P.}~\bibnamefont {Schneider}}, \bibinfo {author}
  {\bibfnamefont {S.}~\bibnamefont {Giglberger}}, \bibinfo {author}
  {\bibfnamefont {J.}~\bibnamefont {Eroms}}, \bibinfo {author} {\bibfnamefont
  {J.~D.}\ \bibnamefont {Boeck}}, \bibinfo {author} {\bibfnamefont
  {G.}~\bibnamefont {Borghs}}, \bibinfo {author} {\bibfnamefont
  {W.}~\bibnamefont {Wegscheider}}, \bibinfo {author} {\bibfnamefont
  {D.}~\bibnamefont {Weiss}}, \ and\ \bibinfo {author} {\bibfnamefont
  {W.}~\bibnamefont {Prettl}},\ }\href {\doibase 10.1103/physrevlett.92.256601}
  {\bibfield  {journal} {\bibinfo  {journal} {Phys. Rev. Lett.}\ }\textbf
  {\bibinfo {volume} {92}} (\bibinfo {year} {2004}),\
  10.1103/physrevlett.92.256601}\BibitemShut {NoStop}%
\bibitem [{\citenamefont {Mu}\ \emph {et~al.}(2016)\citenamefont {Mu},
  \citenamefont {Sullivan},\ and\ \citenamefont {Du}}]{Mu_2016}%
  \BibitemOpen
  \bibfield  {author} {\bibinfo {author} {\bibfnamefont {X.}~\bibnamefont
  {Mu}}, \bibinfo {author} {\bibfnamefont {G.}~\bibnamefont {Sullivan}}, \ and\
  \bibinfo {author} {\bibfnamefont {R.-R.}\ \bibnamefont {Du}},\ }\href
  {\doibase 10.1063/1.4939230} {\bibfield  {journal} {\bibinfo  {journal}
  {Appl. Phys. Lett.}\ }\textbf {\bibinfo {volume} {108}},\ \bibinfo {pages}
  {012101} (\bibinfo {year} {2016})}\BibitemShut {NoStop}%
\bibitem [{foo()}]{footnote_one}%
  \BibitemOpen
  \href@noop {} {}\bibinfo {note} {Note that this g-factor value of -11.5 is
  measured on a slightly different stack with an InAs layer of 11 nm thick. We
  have checked in the simulations that changing the g-factor to -5 or -15 has
  negligible influence on the SdH oscillations.}\BibitemShut {Stop}%
\bibitem [{\citenamefont {Herzog}\ \emph {et~al.}(2017)\citenamefont {Herzog},
  \citenamefont {Hardtdegen}, \citenamefont {Sch\"apers}, \citenamefont
  {Grundler},\ and\ \citenamefont {Wilde}}]{Herzog_2017}%
  \BibitemOpen
  \bibfield  {author} {\bibinfo {author} {\bibfnamefont {F.}~\bibnamefont
  {Herzog}}, \bibinfo {author} {\bibfnamefont {H.}~\bibnamefont {Hardtdegen}},
  \bibinfo {author} {\bibfnamefont {T.}~\bibnamefont {Sch\"apers}}, \bibinfo
  {author} {\bibfnamefont {D.}~\bibnamefont {Grundler}}, \ and\ \bibinfo
  {author} {\bibfnamefont {M.}~\bibnamefont {Wilde}},\ }\href@noop {}
  {\bibfield  {journal} {\bibinfo  {journal} {ArXiv}\ } (\bibinfo {year}
  {2017})},\ \Eprint {http://arxiv.org/abs/1703.07143} {1703.07143}
  \BibitemShut {NoStop}%
\bibitem [{\citenamefont {Bychkov}\ and\ \citenamefont
  {Rashba}(1984{\natexlab{a}})}]{Bychkov_1984a}%
  \BibitemOpen
  \bibfield  {author} {\bibinfo {author} {\bibfnamefont {Y.~A.}\ \bibnamefont
  {Bychkov}}\ and\ \bibinfo {author} {\bibfnamefont {E.~I.}\ \bibnamefont
  {Rashba}},\ }\href@noop {} {\bibfield  {journal} {\bibinfo  {journal}
  {Journal of physics C: Solid state physics}\ }\textbf {\bibinfo {volume}
  {17}},\ \bibinfo {pages} {6039} (\bibinfo {year}
  {1984}{\natexlab{a}})}\BibitemShut {NoStop}%
\bibitem [{\citenamefont {Bychkov}\ and\ \citenamefont
  {Rashba}(1984{\natexlab{b}})}]{Bychkov_1984b}%
  \BibitemOpen
  \bibfield  {author} {\bibinfo {author} {\bibfnamefont {Y.~A.}\ \bibnamefont
  {Bychkov}}\ and\ \bibinfo {author} {\bibfnamefont {E.}~\bibnamefont
  {Rashba}},\ }\href@noop {} {\bibfield  {journal} {\bibinfo  {journal} {JETP
  lett}\ }\textbf {\bibinfo {volume} {39}},\ \bibinfo {pages} {78} (\bibinfo
  {year} {1984}{\natexlab{b}})}\BibitemShut {NoStop}%
\bibitem [{Note1()}]{Note1}%
  \BibitemOpen
  \bibinfo {note} {We cannot directly extract the spin-orbit strength from this
  $\Delta n$ by comparing to the single-carrier case, since the effective mass
  in this region is unknown.}\BibitemShut {Stop}%
\bibitem [{\citenamefont {Luo}\ \emph {et~al.}(1990)\citenamefont {Luo},
  \citenamefont {Munekata}, \citenamefont {Fang},\ and\ \citenamefont
  {Stiles}}]{Luo_1990}%
  \BibitemOpen
  \bibfield  {author} {\bibinfo {author} {\bibfnamefont {J.}~\bibnamefont
  {Luo}}, \bibinfo {author} {\bibfnamefont {H.}~\bibnamefont {Munekata}},
  \bibinfo {author} {\bibfnamefont {F.~F.}\ \bibnamefont {Fang}}, \ and\
  \bibinfo {author} {\bibfnamefont {P.~J.}\ \bibnamefont {Stiles}},\ }\href
  {\doibase 10.1103/physrevb.41.7685} {\bibfield  {journal} {\bibinfo
  {journal} {Phys. Rev. B}\ }\textbf {\bibinfo {volume} {41}},\ \bibinfo
  {pages} {7685} (\bibinfo {year} {1990})}\BibitemShut {NoStop}%
\end{thebibliography}%


\begin{thebibliography}{2}%
	\makeatletter
	\providecommand \@ifxundefined [1]{%
		\@ifx{#1\undefined}
	}%
	\providecommand \@ifnum [1]{%
		\ifnum #1\expandafter \@firstoftwo
		\else \expandafter \@secondoftwo
		\fi
	}%
	\providecommand \@ifx [1]{%
		\ifx #1\expandafter \@firstoftwo
		\else \expandafter \@secondoftwo
		\fi
	}%
	\providecommand \natexlab [1]{#1}%
	\providecommand \enquote  [1]{``#1''}%
	\providecommand \bibnamefont  [1]{#1}%
	\providecommand \bibfnamefont [1]{#1}%
	\providecommand \citenamefont [1]{#1}%
	\providecommand \href@noop [0]{\@secondoftwo}%
	\providecommand \href [0]{\begingroup \@sanitize@url \@href}%
	\providecommand \@href[1]{\@@startlink{#1}\@@href}%
	\providecommand \@@href[1]{\endgroup#1\@@endlink}%
	\providecommand \@sanitize@url [0]{\catcode `\\12\catcode `\$12\catcode
		`\&12\catcode `\#12\catcode `\^12\catcode `\_12\catcode `\%12\relax}%
	\providecommand \@@startlink[1]{}%
	\providecommand \@@endlink[0]{}%
	\providecommand \url  [0]{\begingroup\@sanitize@url \@url }%
	\providecommand \@url [1]{\endgroup\@href {#1}{\urlprefix }}%
	\providecommand \urlprefix  [0]{URL }%
	\providecommand \Eprint [0]{\href }%
	\providecommand \doibase [0]{http://dx.doi.org/}%
	\providecommand \selectlanguage [0]{\@gobble}%
	\providecommand \bibinfo  [0]{\@secondoftwo}%
	\providecommand \bibfield  [0]{\@secondoftwo}%
	\providecommand \translation [1]{[#1]}%
	\providecommand \BibitemOpen [0]{}%
	\providecommand \bibitemStop [0]{}%
	\providecommand \bibitemNoStop [0]{.\EOS\space}%
	\providecommand \EOS [0]{\spacefactor3000\relax}%
	\providecommand \BibitemShut  [1]{\csname bibitem#1\endcsname}%
	\let\auto@bib@innerbib\@empty
	\bibitem [{\citenamefont {Luo}\ \emph {et~al.}(1990)\citenamefont {Luo},
		\citenamefont {Munekata}, \citenamefont {Fang},\ and\ \citenamefont
		{Stiles}}]{Luo_1990_supp}%
	\BibitemOpen
	\bibfield  {author} {\bibinfo {author} {\bibfnamefont {J.}~\bibnamefont
			{Luo}}, \bibinfo {author} {\bibfnamefont {H.}~\bibnamefont {Munekata}},
		\bibinfo {author} {\bibfnamefont {F.~F.}\ \bibnamefont {Fang}}, \ and\
		\bibinfo {author} {\bibfnamefont {P.~J.}\ \bibnamefont {Stiles}},\ }\href
	{\doibase 10.1103/physrevb.41.7685} {\bibfield  {journal} {\bibinfo
			{journal} {Phys. Rev. B}\ }\textbf {\bibinfo {volume} {41}},\ \bibinfo
		{pages} {7685} (\bibinfo {year} {1990})}\BibitemShut {NoStop}%
	\bibitem [{\citenamefont {Winkler}(2003)}]{Winkler_2003_supp}%
	\BibitemOpen
	\bibfield  {author} {\bibinfo {author} {\bibfnamefont {R.}~\bibnamefont
			{Winkler}},\ }\href
	{http://www.ebook.de/de/product/2638263/roland_winkler_spin_orbit_coupling_effects_in_two_dimensional_electron_and_hole_systems.html}
	{\emph {\bibinfo {title} {Spin-orbit Coupling Effects in Two-dimensional
				Electron and Hole Systems}}}\ (\bibinfo  {publisher} {Springer-Verlag GmbH},\
	\bibinfo {year} {2003})\BibitemShut {NoStop}%
\end{thebibliography}

\onecolumngrid
\linespread{1.5}

\setcounter{figure}{0}

\makeatletter 
\renewcommand{\thefigure}{\normalsize\textbf{S\arabic{figure}}}
\renewcommand{\thetable}{\normalsize \textbf{S\arabic{table}}}
\makeatother

\renewcommand\figurename{\normalsize \textbf{Figure}}
\renewcommand\tablename{\normalsize\textbf{Table}}

\newpage

\begin{center}
\qquad \\[0.1cm] \LARGE{ \huge Supplementary Information \\[5mm] \LARGE Spin-orbit interaction in a dual gated InAs/GaSb quantum well}\\[1cm]

\linespread{1.5}
\large \noindent Arjan J. A. Beukman, Folkert K. de Vries, Jasper van Veen, Rafal Skolasinski, Micheal Wimmer, Fanming Qu, David T. de Vries, Binh-Minh Nguyen, Wei Yi, Andrey A. Kiselev, Marko Sokolich, Michael J. Manfra, Fabrizio Nichele, Charles M. Marcus, and Leo P. Kouwenhoven\\[2cm]
\end{center}

\normalsize
\clearpage
\section{Fourier transforms}
\noindent The Fourier transforms in this manuscript are obtained using the method described here. Starting from a magnetoresistance curve, first a magnetic field range is chosen. The lower bound is fixed at 0.15 T. The upper bound is chosen such that the interval ends at 40\% of a beat maximum. Truncating the signal in this way causes minimal deviation from the true frequency components. Next, the background resistance is estimated using a 6$^{\te{th}}$ order polynomial fit, which subsequently is subtracted from the signal. The remaining signal is interpolated on a uniform grid in $1/B$ and padded with zeros on both sides. No extra window function is applied. A fast Fourier transform converts the signal to the frequency domain $R(\omega)$ and the power spectrum is obtained using $P(\omega) = R(\omega) \times R^*(\omega)$. All Fourier transforms are normalized such that the maximum is 0.8 a.u.

\section{Number of oscillations in a beat}
\begin{figure}[h]
	\centering
	\includegraphics[width=86mm]{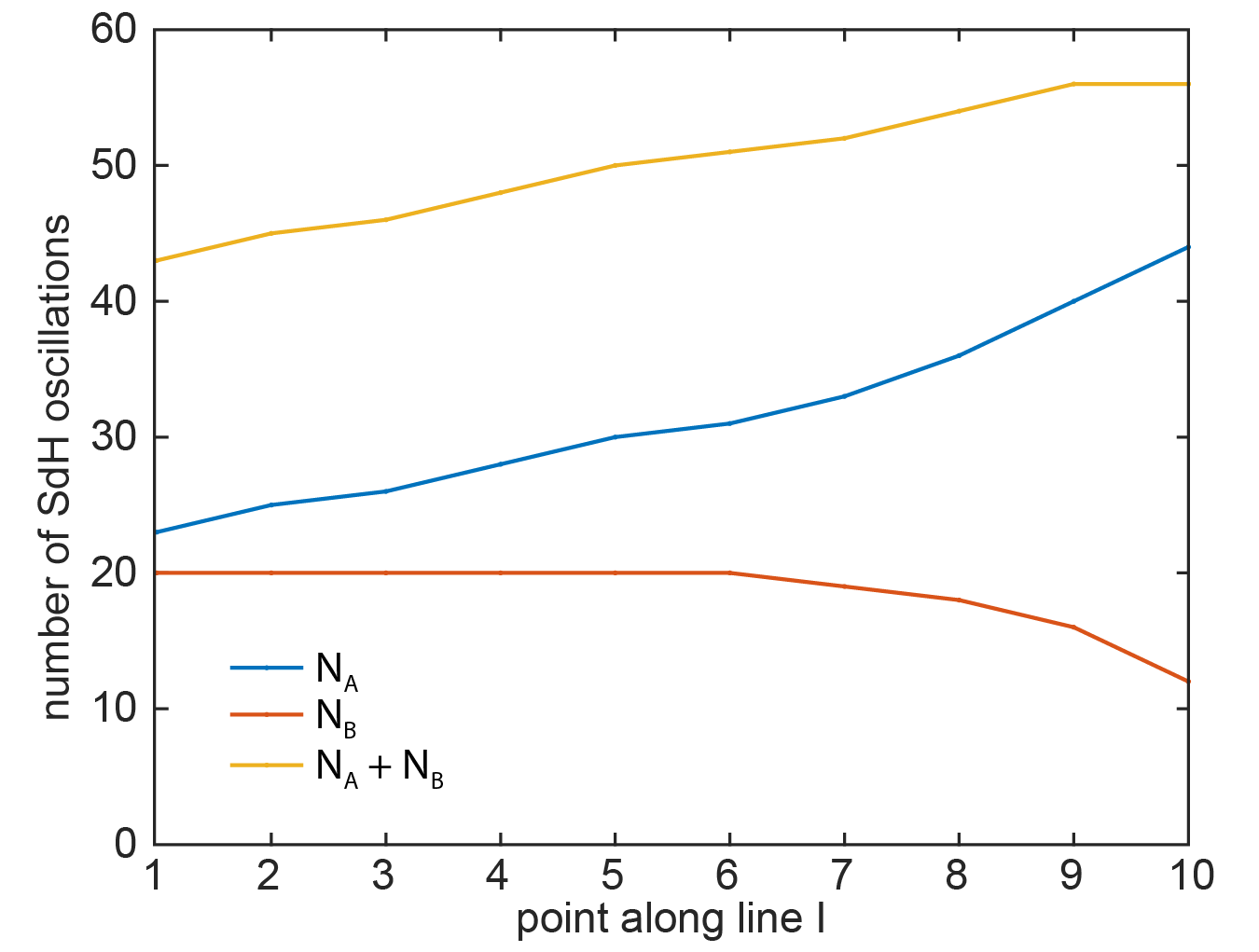}
	\caption{Number of SdH oscillations in beat A ($N_A$), beat B ($N_B$) and the sum ($N_A + N_B$) for each trace in Fig. 2a of the main text. The two trends discussed in the main text are clearly visible here. First, moving from point 1 to 10 the asymmetry $r = {(N_A - N_B)}/{(N_A + N_B)}$ increases. Second, the total number of oscillations $N_A+N_B$ increases.}
	\label{SO:suppl:number_of_oscillations}
\end{figure}
\newpage
\section{Peaks shift upon changing electron density}
\begin{figure}[ht]
	\centering
	\includegraphics[width=110mm]{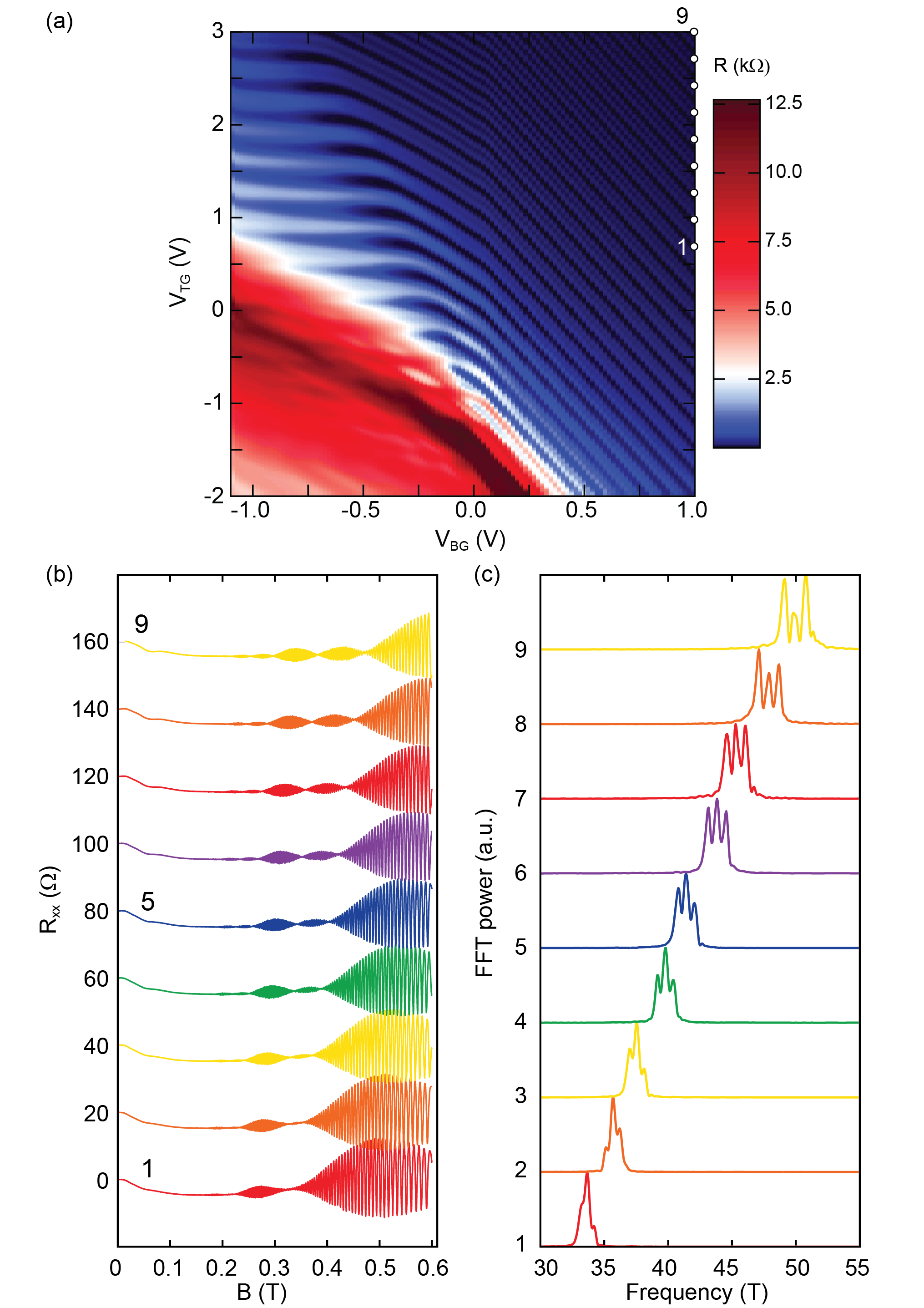}
	\caption{(a) Phase diagram of the InAs/GaSb quantum well at $B_{\perp}= 2$ T showing 9 additional points investigated. Moving from point 1 to 9 several SdH oscillations are crossed, hence the electron density increases from point 1 to 9. (b) Magnetoresistance traces  $R_{xx}(B)$ for points 1 to 9. Traces are offset by 20 $\Omega$. (c) Fourier transforms of the traces in (b), showing that the center frequency shifts together with the outer frequency peaks. This density dependence of the central frequency peak excludes MIS and MAR.}
	\label{SO:suppl:exlude_MIS_MAR}
\end{figure}

\clearpage
\section{Dingle plots of the effective mass}
\begin{figure}[h]
	\centering
	\includegraphics[width=160mm]{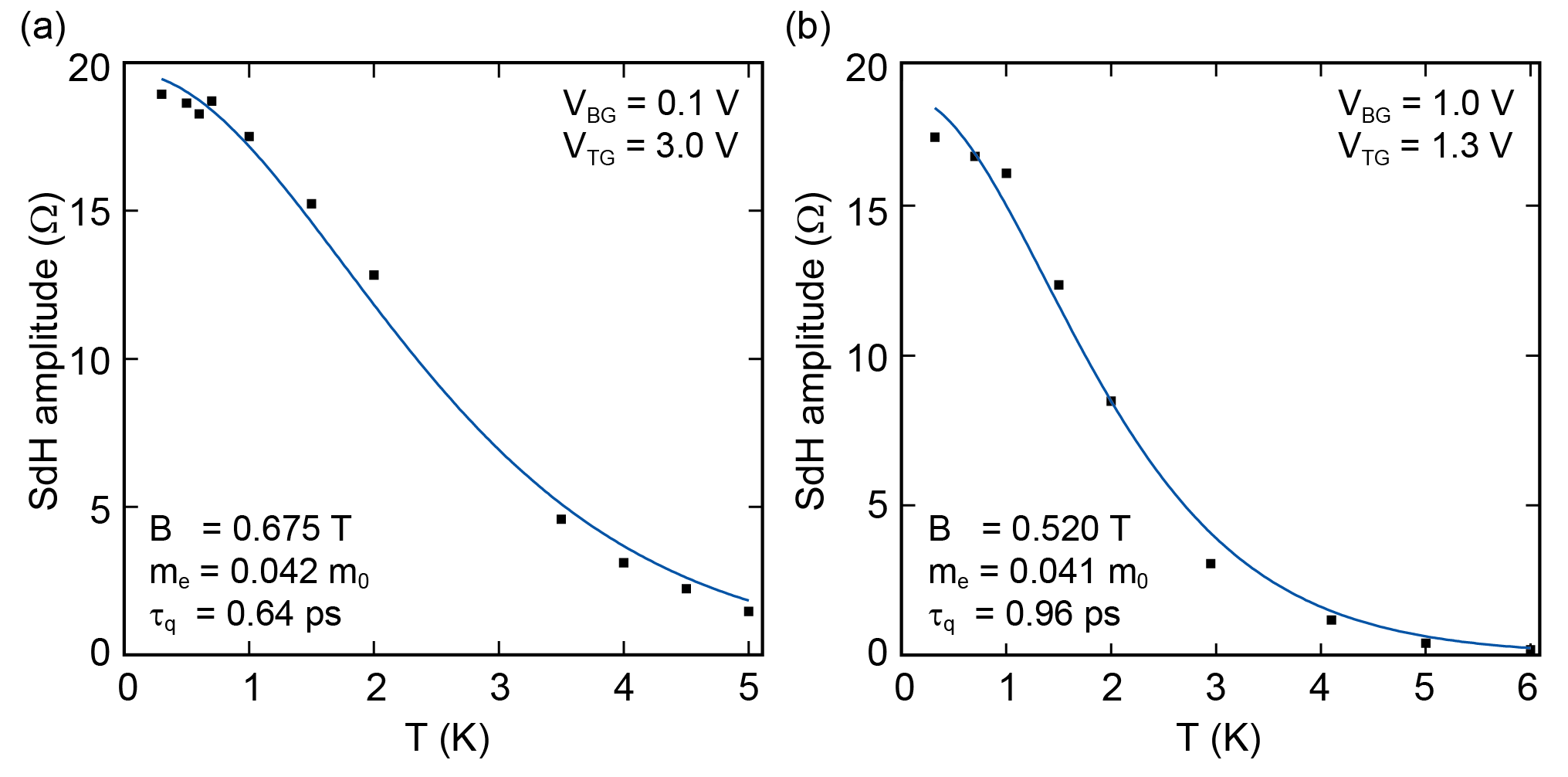}
	\caption{(a,b) Temperature dependence of the SdH oscillations (Dingle plots) measured at point 1 ands 10 on line I, respectively. Effective mass $m^*$, indicated in the figures, is extracted by fitting these data to the Dingle formula (blue curve).}
	\label{SO:suppl:dingle_plots}
\end{figure}

\clearpage
\section{Details on the Landau level simulation}

\noindent This section describes the calculations used to simulate the magnetoresistance traces to extract the Rashba and Dresselhaus coefficients as shown in Fig. 3a of the main text. We closely follow the method presented in Ref. \cite{Luo_1990_supp} and Ch. 4 of Ref. \cite{Winkler_2003_supp}. 

The Hamiltonian in the momentum basis is presented in Eq. 1 of the main text, here repeated for convenience:
\begin{equation}
	\begin{array}{c}
		H = \frac{(\hat{p}_x^2 + \hat{p}_y^2)}{2m^*} \sigma_0 + \alpha(\hat{p}_y \sigma_x-\hat{p}_x\sigma_y)/\hbar + \beta(\hat{p}_x\sigma_x - \hat{p}_y\sigma_y)/\hbar \\
		+ \ \gamma (\hat{p}_y \hat{p}_x \hat{p}_y \sigma_x - \hat{p}_x \hat{p}_y \hat{p}_x \sigma_y)/\hbar^3 + \frac{1}{2}g\mu_B B_z \sigma_z
	\end{array}
\end{equation}
For the perpendicular magnetic field $B=(0,0,B_z)$, the symmetric gauge $\mathbf{A}(x,y) = \frac{B_z}{2} (-y,x,0)$ is used. The canonical momentum can be written as
\begin{equation}
\hat{p} = -i\hbar \nabla + e\mathbf{A}.
\end{equation}
Raising and lowering operators are defined as
\begin{equation}
\begin{array}{rr}
a^\dagger =& \displaystyle \frac{\lambda_c}{\sqrt{2}\hbar} \left( \hat{p}_x + i\hat{p}_y \right),\\[4mm]
a =&   \displaystyle \frac{\lambda_c}{\sqrt{2}\hbar} \left( \hat{p}_x - i\hat{p}_y \right),
\end{array}
\end{equation}
where $\lambda_c = \sqrt{\hbar/eB}$ is the magnetic length. The raising operators act on the Landau levels, i.e. $a^\dagger |n,\uparrow \rangle = \sqrt{n+1} \,|n+1,\uparrow \rangle$. The momentum operators are rewritten in the raising and lowering operators, which are then substituted into the Hamiltonian. We take a basis of $N=400$ Landau levels in order to capture magnetic fields $\gtrsim 0.1$ T for the electron density $n_s \simeq 17.6 \cdot 10^{15}$ m$^{-2}$. Solving the Hamiltonian results in the Landau level energies at a particular magnetic field $E(n,B_z)$.

Following Luo et al. \cite{Luo_1990_supp} the conductance is written as:
\begin{equation}
\sigma_{xx} = \frac{e^2}{\pi^2 \hbar} \sum_{n, \uparrow \downarrow} \left(n \pm \frac{1}{2}\right) \exp \left( \displaystyle -\frac{(E_f-E_{n,\uparrow \downarrow})^2}{\Gamma^2} \right).
\end{equation}
We assume a fixed Fermi energy at $E_f = (\pi \hbar^2 n_s)/m^*$. To obtain the resistivity we use the approximation that for quantizing magnetic fields $\left( \sigma_{xy}^2 \gg \sigma_{xx}^2 \right)$ the transverse resistivity $\rho_{xx}$ is given as \cite{Luo_1990_supp}:
\begin{equation}
\rho_{xx} = \sigma_{xx}/(\sigma_{xy}^2 + \sigma_{xx}^2) \approx \sigma_{xx}/\sigma_{xy}^2 \approx \sigma_{xx} \left( B_z/e n_s \right)^2
\end{equation}

\clearpage
\section{SdH traces of the Landau level simulations}
\begin{figure}[h]
	\centering
	\includegraphics[width=90mm]{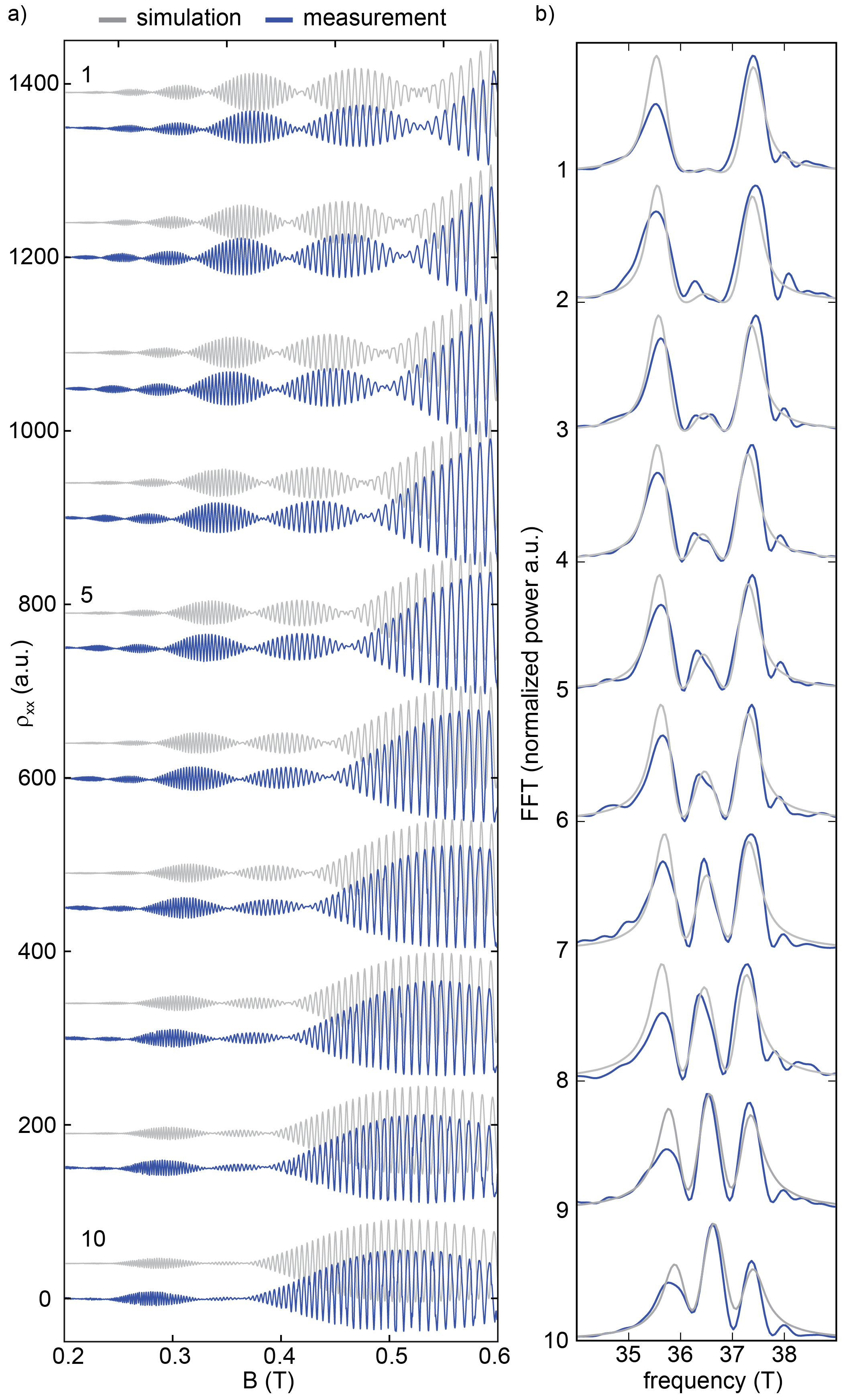}
	\caption{(a)
Measured magnetoresistance traces (blue) for points 1-10 along line I together with the simulated traces (gray). The simulated traces are offset 40 a.u. from the measured traces; the measured traces are offset by 150 a.u. from each other. The value of Rashba spin-orbit interaction strength $\alpha$ is linearly changed from 73 to 53 mV\AA \ going from trace 1 to 10. The linear and cubic Dresselhaus interaction strength are kept constant at $\beta = 28.5$ mV\AA \ and $\gamma = 0$ mV\AA$^3$. (b) Corresponding Fourier transforms of the measured traces (blue) and simulated traces (gray), i.e. a reproduction of Fig. 3c.}
	\label{SO:suppl:LL_simulation_results}
\end{figure}

\clearpage
\section{Fourier transforms in the two-carrier regime}
\begin{figure}[h]
	\centering
	\includegraphics[width=90mm]{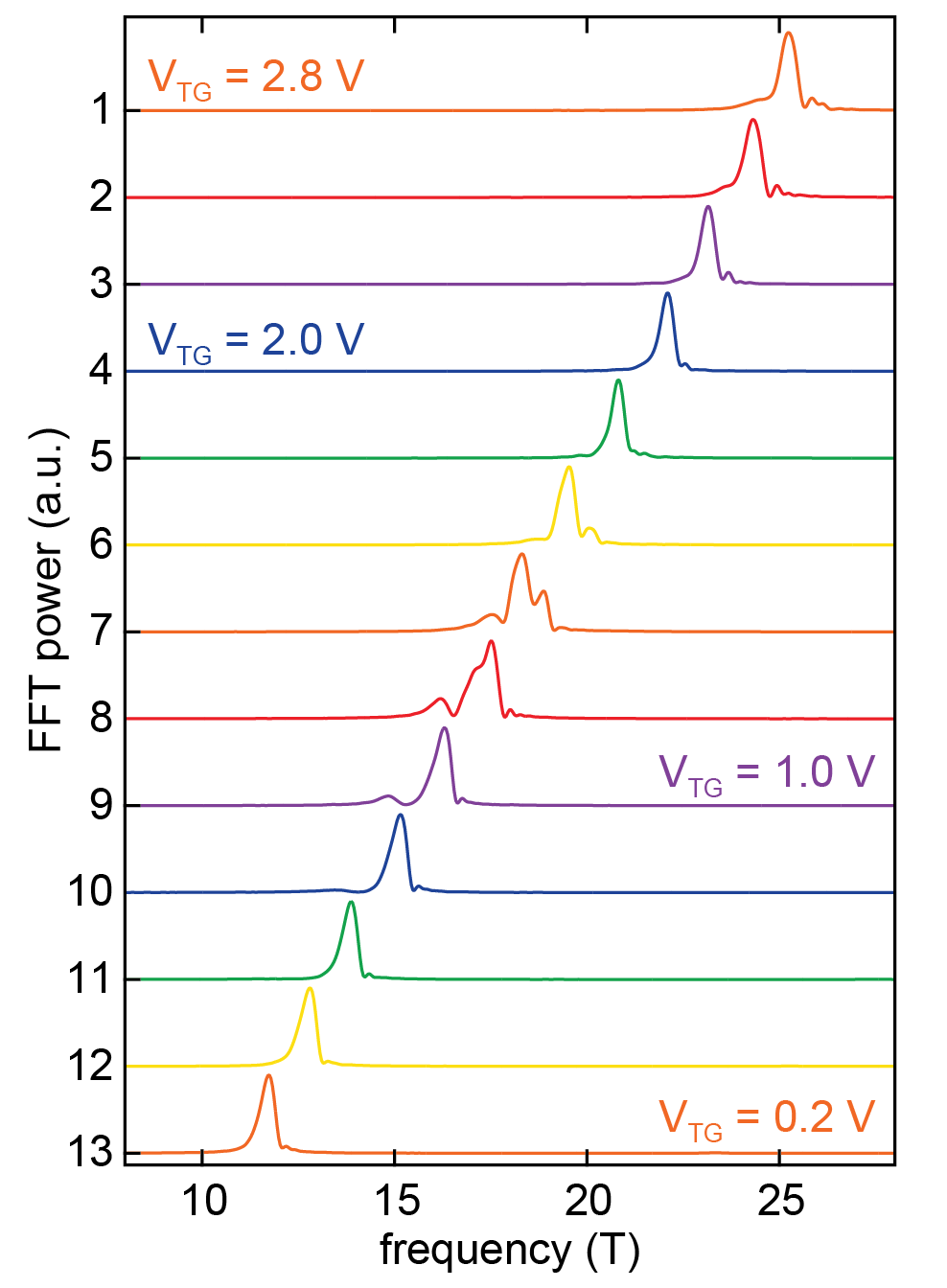}
	\caption{Fourier transforms for the 13 points along line II in the two-carrier regime. All traces are normalized such that the maximum is set to 0.8 a.u.}
	\label{SO:suppl:inverted_FFT}
\end{figure}


\clearpage
\section{Band structure calculations for multiple electric fields}
\begin{figure}[h]
	\centering
	\includegraphics[width=\linewidth]{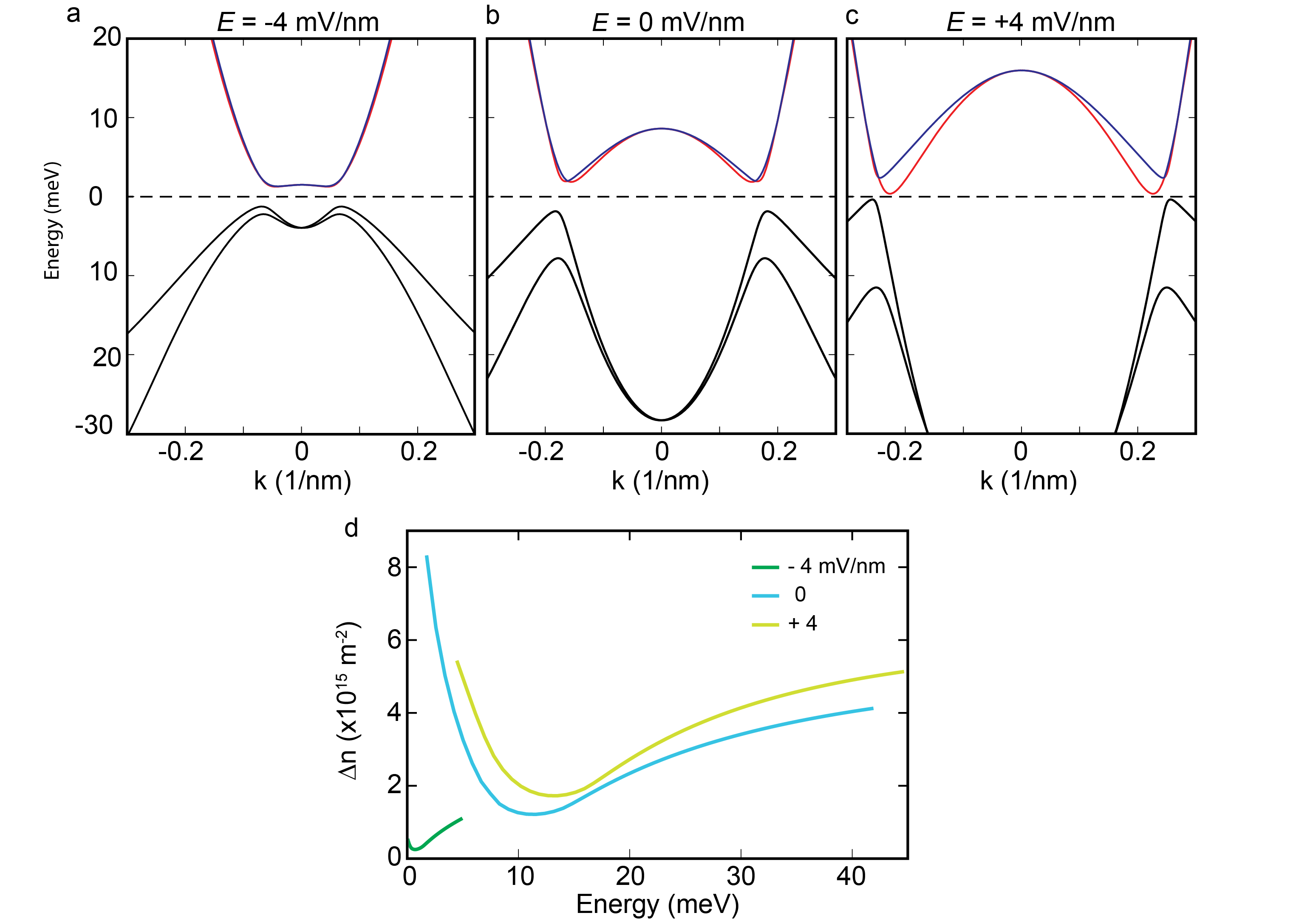}
	\caption{(a-c). Result of band structure calculations along [100] for electric fields $\vec{E} = -4$ mV/nm, 0 mV/nm and $+4$ mV/nm respectively. The spin-split bands of the conduction band are colored red and blue for clarity. (d) The ZFSS extracted from these band structures. For all electric fiels a minimum in spin-splitting is found, making this a robust feature of an inverted band structure.}
	\label{SO:suppl:inverted_FFT}
\end{figure}

\clearpage
\section{Effective mass \& wave function in the quantum well}
\begin{figure}[h]
	\centering
	\includegraphics[width=75mm]{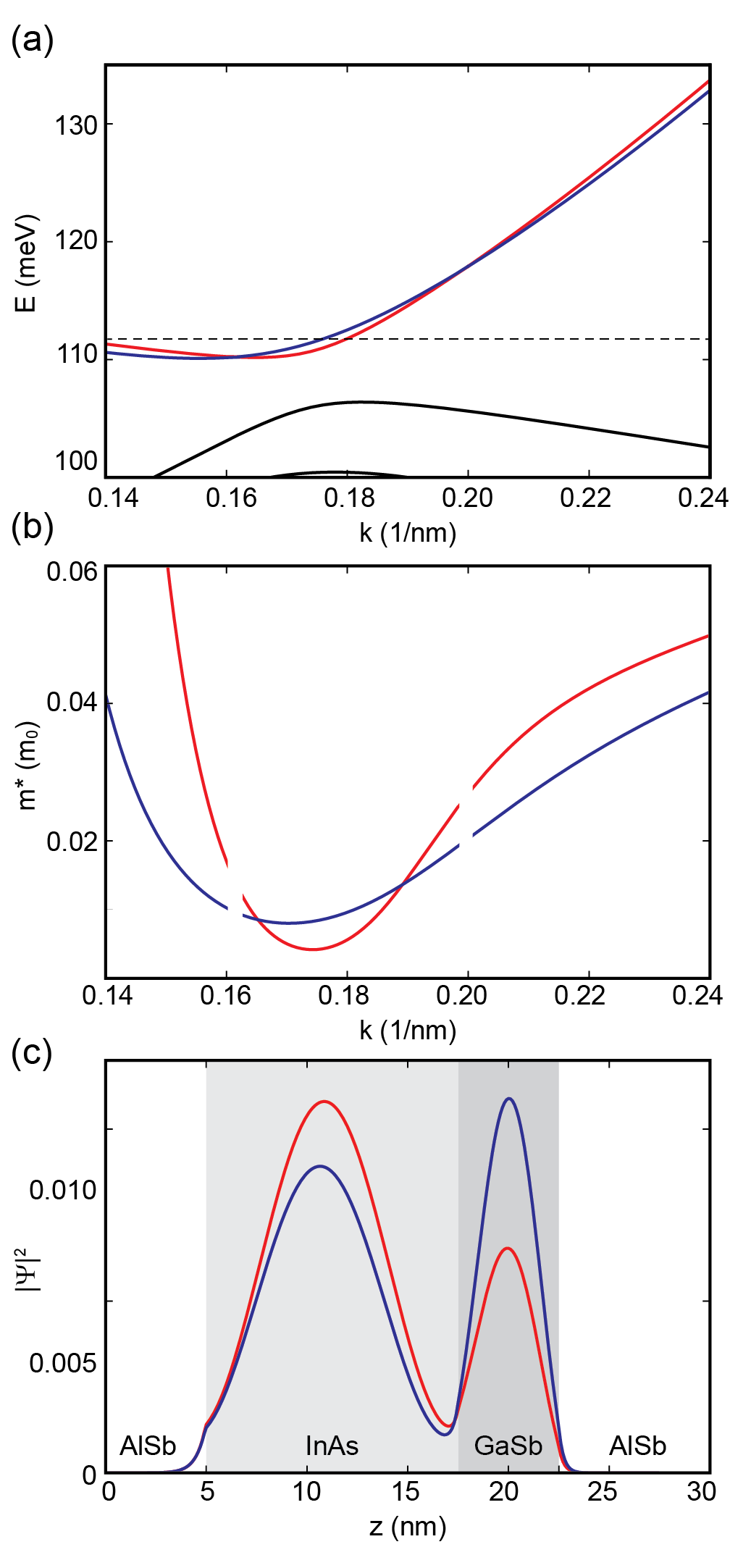}
	\caption{(a) Zoom-in on the calculated band structure (at $\vec{E} = 0$) close to the hybridization gap. Red and blue correspond to different spin species. (b) Effective mass calculated from $m^* = \hbar^2 \left( \partial^2 E/\partial k^2 \right)^{-1} $. The effective mass around $k=0.16$ nm$^{-1}$ and $k=0.20$ nm$^{-1}$ is omitted due to numerical artifacts. (c) Simulated normalized wave functions at the energy indicated by the dashed line in (a). It is clear that the distribution of the wave function is spin dependent, which could result in a spin dependent scattering.}
	\label{SO:suppl:effective_mass}
\end{figure}

\end{document}